\input harvmac
\let\includefigures=\iftrue
\let\useblackboard=\iftrue
\newfam\black

\includefigures
\message{If you do not have epsf.tex (to include figures),}
\message{change the option at the top of the tex file.}
\input epsf
\def\figin{\epsfcheck\figin}\def\figins{\epsfcheck\figins}
\def\epsfcheck{\ifx\epsfbox\UnDeFiNeD
\message{(NO epsf.tex, FIGURES WILL BE IGNORED)}
\gdef\figin##1{\vskip2in}\gdef\figins##1{\hskip.5in}
\else\message{(FIGURES WILL BE INCLUDED)}%
\gdef\figin##1{##1}\gdef\figins##1{##1}\fi}
\def\DefWarn#1{}
\def\figinsert{\goodbreak\midinsert}
\def\ifig#1#2#3{\DefWarn#1\xdef#1{fig.~\the\figno}
\writedef{#1\leftbracket fig.\noexpand~\the\figno}%
\figinsert\figin{\centerline{#3}}\medskip\centerline{\vbox{
\baselineskip12pt\advance\hsize by -1truein
\noindent\footnotefont{\bf Fig.~\the\figno:} #2}}
\endinsert\global\advance\figno by1}
\else
\def\ifig#1#2#3{\xdef#1{fig.~\the\figno}
\writedef{#1\leftbracket fig.\noexpand~\the\figno}%
\global\advance\figno by1} \fi
%
\def\wt{\widetilde}

\def\id{{1 \kern-.28em {\rm l}}}
\def\N{{\cal N}}

\def\K3{{\bf K3}}
\def\journal#1&#2(#3){\unskip, \sl #1\ \bf #2 \rm(19#3) }
\def\andjournal#1&#2(#3){\sl #1~\bf #2 \rm (19#3) }

\def\bar{\overline}
\def\hat{\widehat}
\def\ie{{\it i.e.}}
\def\eg{{\it e.g.}}

\def\tilde{\widetilde}

\def\frac#1#2{{#1\over#2}}

\def\half{\frac12}

\def\d{\partial}

\def\inbar{\,\vrule height1.5ex width.4pt depth0pt}
\def\IC{\relax\hbox{$\inbar\kern-.3em{\rm C}$}}
\def\IR{\relax{\rm I\kern-.18em R}}
\def\IP{\relax{\rm I\kern-.18em P}}

%
%

%
\catcode`\@=11
\def\slash#1{\mathord{\mathpalette\c@ncel{#1}}}
\overfullrule=0pt

\def\EE{{\cal E}}

\def\OO{{\cal O}}

\def\SS{{\cal S}}

\def\underrel#1\over#2{\mathrel{\mathop{\kern\z@#1}\limits_{#2}}}

\catcode`\@=12


%

\def\exp{{\rm exp}}


\def\vbar{{\bar v}}
\def\wbar{{\overline w}}

\lref\ElitzurFH{
  S.~Elitzur, A.~Giveon and D.~Kutasov,
  ``Branes and N = 1 duality in string theory,''
  Phys.\ Lett.\  B {\bf 400}, 269 (1997)
  [arXiv:hep-th/9702014].
}

\lref\ElitzurHC{
  S.~Elitzur, A.~Giveon, D.~Kutasov, E.~Rabinovici and A.~Schwimmer,
  ``Brane dynamics and N = 1 supersymmetric gauge theory,''
  Nucl.\ Phys.\  B {\bf 505}, 202 (1997)
  [arXiv:hep-th/9704104].
}

\lref\GiveonSR{
  A.~Giveon and D.~Kutasov,
  ``Brane dynamics and gauge theory,''
  Rev.\ Mod.\ Phys.\  {\bf 71}, 983 (1999)
  [arXiv:hep-th/9802067].
}

\lref\EssigKZ{
  R.~Essig, J.~F.~Fortin, K.~Sinha, G.~Torroba and M.~J.~Strassler,
  ``Metastable supersymmetry breaking and multitrace deformations of SQCD,''
  arXiv:0812.3213 [hep-th].
}

\lref\IntriligatorDD{
  K.~A.~Intriligator, N.~Seiberg and D.~Shih,
  ``Dynamical SUSY breaking in meta-stable vacua,''
  JHEP {\bf 0604}, 021 (2006)
  [arXiv:hep-th/0602239].
}

\lref\GiveonEW{
  A.~Giveon and D.~Kutasov,
  ``Stable and Metastable Vacua in Brane Constructions of SQCD,''
  JHEP {\bf 0802}, 038 (2008)
  [arXiv:0710.1833 [hep-th]].
}

\lref\GiveonEF{
  A.~Giveon and D.~Kutasov,
  ``Stable and Metastable Vacua in SQCD,''
  Nucl.\ Phys.\  B {\bf 796}, 25 (2008)
  [arXiv:0710.0894 [hep-th]].
}

\lref\GiveonFK{
  A.~Giveon and D.~Kutasov,
  ``Gauge symmetry and supersymmetry breaking from intersecting branes,''
  Nucl.\ Phys.\  B {\bf 778}, 129 (2007)
  [arXiv:hep-th/0703135].
}

\lref\NambuTP{Y.~Nambu and G.~Jona-Lasinio,
``Dynamical Model Of Elementary Particles Based On An Analogy With
Superconductivity. I,''Phys.\ Rev.\  {\bf 122}, 345 (1961).
}

\lref\AntonyanVW{
  E.~Antonyan, J.~A.~Harvey, S.~Jensen and D.~Kutasov,
  ``NJL and QCD from string theory,''
  arXiv:hep-th/0604017.
}

\lref\CallanAT{
  C.~G.~Callan, J.~A.~Harvey and A.~Strominger,
  ``Supersymmetric string solitons,''
  arXiv:hep-th/9112030.
}

\lref\MarolfVF{
  D.~Marolf, L.~Martucci and P.~J.~Silva,
  ``Actions and fermionic symmetries for D-branes in bosonic backgrounds,''
  JHEP {\bf 0307}, 019 (2003)
  [arXiv:hep-th/0306066].
}

\lref\WessCP{
  J.~Wess and J.~Bagger,
  ``Supersymmetry and supergravity,''
{\it  Princeton, USA: Univ. Pr. (1992) 259 p.}
}

\lref\BeasleyDC{
  C.~Beasley, J.~J.~Heckman and C.~Vafa,
  JHEP {\bf 0901}, 058 (2009)
  [arXiv:0802.3391 [hep-th]].
}

\lref\BeasleyKW{
  C.~Beasley, J.~J.~Heckman and C.~Vafa,
  JHEP {\bf 0901}, 059 (2009)
  [arXiv:0806.0102 [hep-th]].
}

\lref\OoguriBG{
  H.~Ooguri and Y.~Ookouchi,
  ``Meta-stable supersymmetry breaking vacua on intersecting branes,''
  Phys.\ Lett.\  B {\bf 641}, 323 (2006)
  [arXiv:hep-th/0607183].
}

\lref\FrancoHT{
  S.~Franco, I.~Garcia-Etxebarria and A.~M.~Uranga,
  ``Non-supersymmetric meta-stable vacua from brane configurations,''
  JHEP {\bf 0701}, 085 (2007)
  [arXiv:hep-th/0607218].
}

\lref\BenaRG{
  I.~Bena, E.~Gorbatov, S.~Hellerman, N.~Seiberg and D.~Shih,
  ``A note on (meta)stable brane configurations in MQCD,''
  JHEP {\bf 0611}, 088 (2006)
  [arXiv:hep-th/0608157].
}

\lref\PolchinskiRR{
  J.~Polchinski,
  ``String theory. Vol. 2: Superstring theory and beyond,''
{\it  Cambridge, UK: Univ. Pr. (1998) 531 p.}
}

\lref\GiudiceBP{
  G.~F.~Giudice and R.~Rattazzi,
  ``Theories with gauge-mediated supersymmetry breaking,''
  Phys.\ Rept.\  {\bf 322}, 419 (1999)
  [arXiv:hep-ph/9801271].
}

\lref\AganagicPE{
  M.~Aganagic, C.~Popescu and J.~H.~Schwarz,
  ``D-brane actions with local kappa symmetry,''
  Phys.\ Lett.\  B {\bf 393}, 311 (1997)
  [arXiv:hep-th/9610249].
}

\lref\BergshoeffTU{
  E.~Bergshoeff and P.~K.~Townsend,
  ``Super D-branes,''
  Nucl.\ Phys.\  B {\bf 490}, 145 (1997)
  [arXiv:hep-th/9611173].
}

\lref\GiveonUR{
  A.~Giveon, D.~Kutasov, J.~McOrist and A.~B.~Royston,
  ``D-Term Supersymmetry Breaking from Branes,''
  arXiv:0904.0459 [hep-th].
}

\lref\CampbellHD{
  B.~A.~Campbell, S.~Davidson and K.~A.~Olive,
  ``Inflation, neutrino baryogenesis, and (S)neutrino induced baryogenesis,''
  Nucl.\ Phys.\  B {\bf 399}, 111 (1993)
  [arXiv:hep-ph/9302223].
}

\lref\AffleckFY{
  I.~Affleck and M.~Dine,
  ``A New Mechanism For Baryogenesis,''
  Nucl.\ Phys.\  B {\bf 249}, 361 (1985).
}

\lref\FalkZT{
  T.~Falk, K.~A.~Olive, L.~Roszkowski, A.~Singh and M.~Srednicki,
  ``Constraints from inflation and reheating on superpartner masses,''
  Phys.\ Lett.\  B {\bf 396}, 50 (1997)
  [arXiv:hep-ph/9611325].
}

\lref\KomargodskiJF{
  Z.~Komargodski and D.~Shih,
  ``Notes on SUSY and R-Symmetry Breaking in Wess-Zumino Models,''
  JHEP {\bf 0904}, 093 (2009)
  [arXiv:0902.0030 [hep-th]].
}

\lref\GiveonZM{
  A.~Giveon, D.~Kutasov and O.~Pelc,
  ``Holography for non-critical superstrings,''
  JHEP {\bf 9910}, 035 (1999)
  [arXiv:hep-th/9907178].
}

\lref\HorowitzCD{
  G.~T.~Horowitz and A.~Strominger,
  ``Black strings and P-branes,''
  Nucl.\ Phys.\  B {\bf 360}, 197 (1991).
}

\lref\MaldacenaCG{
  J.~M.~Maldacena and A.~Strominger,
  ``Semiclassical decay of near-extremal fivebranes,''
  JHEP {\bf 9712}, 008 (1997)
  [arXiv:hep-th/9710014].
}

\lref\AbelCR{
  S.~A.~Abel, C.~S.~Chu, J.~Jaeckel and V.~V.~Khoze,
  ``SUSY breaking by a metastable ground state: Why the early universe
  preferred the non-supersymmetric vacuum,''
  JHEP {\bf 0701}, 089 (2007)
  [arXiv:hep-th/0610334].
}

\lref\FischlerXH{
  W.~Fischler, V.~Kaplunovsky, C.~Krishnan, L.~Mannelli and M.~A.~C.~Torres,
  ``Meta-Stable Supersymmetry Breaking in a Cooling Universe,''
  JHEP {\bf 0703}, 107 (2007)
  [arXiv:hep-th/0611018].
}

\lref\CraigKX{
  N.~J.~Craig, P.~J.~Fox and J.~G.~Wacker,
  ``Reheating metastable O'Raifeartaigh models,''
  Phys.\ Rev.\  D {\bf 75}, 085006 (2007)
  [arXiv:hep-th/0611006].
}

\lref\AbelMY{
  S.~A.~Abel, J.~Jaeckel and V.~V.~Khoze,
  ``Why the early universe preferred the non-supersymmetric vacuum. II,''
  JHEP {\bf 0701}, 015 (2007)
  [arXiv:hep-th/0611130].
}

\lref\DolanQD{
  L.~Dolan and R.~Jackiw,
  ``Symmetry Behavior At Finite Temperature,''
  Phys.\ Rev.\  D {\bf 9}, 3320 (1974).
}

\lref\ElitzurFH{
  S.~Elitzur, A.~Giveon and D.~Kutasov,
  ``Branes and N = 1 duality in string theory,''
  Phys.\ Lett.\  B {\bf 400}, 269 (1997)
  [arXiv:hep-th/9702014].
}

\lref\GiveonBV{
  A.~Giveon, D.~Kutasov and O.~Lunin,
  ``Spontaneous SUSY Breaking in Various Dimensions,''
  arXiv:0904.2175 [hep-th].
}

\lref\SeibergPQ{
  N.~Seiberg,
  ``Electric - magnetic duality in supersymmetric nonAbelian gauge theories,''
  Nucl.\ Phys.\  B {\bf 435}, 129 (1995)
  [arXiv:hep-th/9411149].
}

\lref\KoschadeQU{
  D.~Koschade, M.~McGarrie and S.~Thomas,
  ``Direct Mediation and Metastable Supersymmetry Breaking for SO(10),''
  arXiv:0909.0233 [hep-ph].
}

\lref\GiveonYU{
  A.~Giveon, A.~Katz and Z.~Komargodski,
  ``Uplifted Metastable Vacua and Gauge Mediation in SQCD,''
  JHEP {\bf 0907}, 099 (2009)
  [arXiv:0905.3387 [hep-th]].
}

\lref\sjrey{S. J. Rey, unpublished.}

\Title{}
{\vbox{\centerline{Dynamical Vacuum Selection in String Theory}
\bigskip
\centerline{}
}}
\bigskip

\centerline{\it  David Kutasov$^1$, Oleg Lunin$^{1,2}$, Jock McOrist$^{1,3}$ and Andrew B. Royston$^{1}$}
\bigskip
\smallskip
\centerline{${}^{1}$EFI and Department of Physics, University of
Chicago} \centerline{5640 S. Ellis Av., Chicago, IL 60637, USA }
\smallskip
\centerline{${}^{2}$Department of Physics and Astronomy,
University of Kentucky} \centerline{Lexington, Kentucky 40506-0055, USA}
\smallskip
\centerline{${}^{3}$DAMTP, Centre for Mathematical Sciences} 
\centerline{ Wilberforce Road, Cambridge, CB3 OWA, UK}
\smallskip

\vglue .3cm

\bigskip

\let\includefigures=\iftrue
\bigskip
\noindent
We study a system of $D$-branes localized near an intersection of Neveu-Schwarz fivebranes,
that is known to exhibit a landscape of supersymmetric and (metastable) supersymmetry
breaking vacua. We show that early universe cosmology drives it to a particular
long-lived supersymmetry breaking ground state.

\bigskip

\Date{}

\newsec{Introduction}

String theory is believed to give rise to a rich landscape of supersymmetric
and non-supersymmetric vacua. A natural question is whether early
universe cosmology can provide a dynamical selection mechanism among these
vacua.  The answer is unknown in general, both because the full structure of
the landscape is at present not well understood, and since real time dynamics
in string theory is often hard to study.

To gain insight into this issue, it might be useful to consider simplified versions of
the landscape, which are expected to play a role in the full problem, but are more
amenable to a detailed analysis. The purpose of this paper is to discuss an example
of a ``mini landscape'' of this sort, focusing on the question of dynamical vacuum selection.

We will consider a system of Neveu-Schwarz fivebranes connected by Dirichlet
fourbranes,  which is known to be useful for realizing supersymmetric gauge theories
in string theory \GiveonSR, and has been used in the last few years to study spontaneous
supersymmetry breaking \refs{\OoguriBG\FrancoHT\BenaRG-\GiveonFK}. In a particular
region of its parameter space, the low energy dynamics of this brane system reduces to a
Wess-Zumino (WZ) model which was analyzed in \IntriligatorDD. In another, one has to take
into account classical string effects, which give rise to non-renormalizable terms in the  low
energy effective Lagrangian,  suppressed by a relatively low UV scale \refs{\GiveonFK,\GiveonUR}.
We will refer to the two regions as the field theory and string theory regimes, respectively.

Interestingly, the vacuum structure found in \IntriligatorDD\ in the field theory regime is very
similar to that obtained in the string theory one in \refs{\GiveonFK,\GiveonUR}. In both cases
one finds a metastable supersymmetry breaking ground state, located at the origin of an
approximate moduli space, and separated by a wide potential barrier from a supersymmetric
ground state. The effects that lift the moduli space are different in the two regions. In the field
theory regime they are due to a one loop contribution to the  effective potential of the low energy
WZ model, while in the string theory regime their origin is the classical gravitational interaction
between $D$-branes and $NS5$-branes.

It is natural to ask whether early universe dynamics drives the system to the true (supersymmetric)
vacuum, or to the metastable (supersymmetry breaking) one. In the field theory regime this question
was addressed in \refs{\AbelCR\CraigKX\FischlerXH-\AbelMY}. These authors  analyzed the structure
of the finite temperature effective potential and argued that, under mild assumptions, the system ends
up in the metastable vacuum.

In this paper we will study this question in the string theory regime. We will see that  early universe
dynamics drives the system to the metastable state, like in the field theory regime, but the details are
different. Furthermore, in string theory one encounters an important issue that does not arise in the
field theory discussion. In studying the early universe dynamics in field theory, the couplings
in the Lagrangian are taken to be fixed (\ie\ time independent).\foot{Except for the usual RG
evolution with the scale.}   In string theory, these couplings are parameters of the brane
configuration, and may depend  on time. We find that the time evolution of these parameters
leads in general to a singular limit of the ISS brane configuration.

Our main purpose in this paper is to study the generalized ISS model which, as we will see,
does not suffer from the above problem. The vacuum structure of this model was analyzed in
\GiveonEW\ in the string theory regime  and in  \refs{\GiveonEF,\EssigKZ} in the field theory one.
It explicitly breaks R-symmetry, and thus is promising for phenomenological applications. The
field theory analysis leads in this case to a number of supersymmetric and metastable
non-supersymmetric vacua, which are generalizations of those found in \IntriligatorDD. In the
string regime one finds many additional metastable vacua, which are qualitatively different
from those of \IntriligatorDD\ and potentially interesting for phenomenology. We will see that early
universe dynamics provides a selection mechanism among the stable and metastable vacua of
the generalized ISS brane system.

The plan of this paper is the following. We start in section 2 by reviewing some field theoretic
results. We describe the ISS model, its R-symmetry breaking generalization, and the results of
\refs{\AbelCR\CraigKX\FischlerXH-\AbelMY} on the early universe evolution of this model. In section
3 we review the brane construction of the (generalized) ISS model and the role of the various
parameters that appear in it. In section 4 we study the dynamics of the brane systems of section
3 in typical excited states that are expected to play a role in the early universe. We argue that the
main effect of the excitation is to turn the Neveu-Schwarz fivebranes into non-extremal black branes
with non-zero horizon radii. The light translational modes on the $D$-branes are excited as well.
Thus, the study of early universe dynamics of the brane systems in the classical string theory
regime amounts to analyzing the dynamics of the $D$-branes in the background of non-extremal fivebranes.

Applying these considerations to the ISS brane system, we find that at early times the system is
in a phase with unbroken gauge symmetry. As the fivebranes relax to their extremal limit, the system
undergoes a phase transition to the metastable state. This conclusion is similar to that found in
the field theory regime in \refs{\AbelCR\CraigKX\FischlerXH-\AbelMY}, but the details are different.
We also point out that the attraction of the non-extremal $NS5$-branes to each other naturally
leads the ISS system at late times to a state in which some of the branes coalesce. Hence the
system of \refs{\OoguriBG\FrancoHT\BenaRG-\GiveonFK} is an unnatural endpoint of this
dynamics. We then turn to the generalized ISS brane configuration which, as mentioned above,
does not suffer from this problem. The rich vacuum structure of this system makes the problem of
vacuum selection particularly interesting. We show that the dynamics of the $D$-branes in the background
of non-extremal fivebranes drives the system to a specific non-supersymmetric ground state.

In section 5 we comment on our results  and possible extensions. Two appendices contain the
details of some of the calculations.

\newsec{Review of field theory results}
In this section, we briefly review the ISS model \IntriligatorDD, its finite temperature extension
\refs{\AbelCR\CraigKX\FischlerXH-\AbelMY}, and the generalization discussed in
\refs{\GiveonEF,\EssigKZ}. The main purpose of this discussion is to set the stage for the
subsequent analysis.

\subsec{The ISS model}

Consider $\N=1$ supersymmetric QCD (SQCD)  with gauge group $SU(N_c)$ and $N_f$ flavors of
(anti) fundamentals $Q_i$, $\tilde Q^i$, $i=1,2,\cdots, N_f$. This (``electric'') theory is believed to be
Seiberg dual \SeibergPQ\ to a ``magnetic'' gauge theory, with gauge group $SU(N_f-N_c)$, $N_f$
fundamentals $q^i$, $\tilde q_i$, and a gauge singlet meson field $\Phi^i_j$, which is coupled to the
magnetic quarks $q$, $\tilde q$ via the superpotential $W_m=hq\Phi\tilde q$.

The two theories are expected to be equivalent in the infrared, \ie\ well below their dynamically
generated scales.  When one of them is weakly coupled at long distances, the other is strongly
coupled. In the range $N_c<N_f<3N_c/2$, where the analysis of \IntriligatorDD\ takes place, the
electric theory is strongly coupled in the infrared, while the magnetic one is infrared free. In this
regime, the electric theory can be viewed as providing a UV completion of the non-asymptotically
free magnetic theory.

To study spontaneous supersymmetry breaking, one needs to turn on a small mass for the electric quarks
$Q$, $\tilde Q$. In the magnetic theory, this corresponds to deforming the superpotential as follows:
\eqn\issw{W_m=h  q^j \Phi^i_j \tilde q_i - h \mu^2 \Phi_i^i~.}
The superfields $\Phi$, $q$, $\tilde q$ are normalized such that their K\"ahler potential is canonical,
up to corrections suppressed by powers of the dynamically generated scales. Thus, the classical
bosonic potential takes the form
\eqn\vveeff{V_0=|h|^2\left(|q\tilde q -\mu^2I_{N_f}|^2+|q\Phi|^2+|\Phi\tilde q|^2\right)~,}
where $I_{N_f}$ is the $N_f\times N_f$ identity matrix. As in the O'Raifeartaigh model,
the potential \vveeff\ cannot be set to zero due to the rank of the matrix $q\wt q$; hence
supersymmetry is broken. The minimum of the potential corresponds (up to global symmetries) to
\eqn\mmqq{q\tilde q=\left(\matrix{ \mu^2 I_{N_f-N_c} & 0 \cr 0 & 0}\right)~,\qquad
\Phi=\left(\matrix{ 0 & 0 \cr 0 & X}\right)~.
}
The $N_c\times N_c$ matrix valued field $X$ has a  flat potential, and  parameterizes a
pseudo-moduli space of  classical vacua. Since supersymmetry is broken, the fate of this
moduli space is sensitive to small corrections to the dynamics. In the regime of interest
for the analysis of \IntriligatorDD, the leading correction is the one loop
Coleman-Weinberg potential, which behaves near the origin like
\eqn\voneloop{V_{1}={\ln4-1\over8\pi^2}(N_f-N_c)|h^2\mu|^2{\rm Tr X^\dagger X}+\cdots~.}
It gives a mass of order $|h^2 \mu|$ to the pseudo-moduli, and stabilizes them at the origin.
Non-perturbative gauge dynamics gives rise to additional, supersymmetric, vacua at
$q=\tilde q=0$ and large $\Phi$,
\eqn\susyvacuum{ \langle\Phi\rangle  = \, \Phi_0=\,
{\mu\over h} \left(\frac{\Lambda}{\mu} \right)^{\frac{ 3N_c- 2N_f}{N_c}}  \, I_{N_f}~,}
where $\Lambda$ is the dynamically generated scale of the magnetic gauge theory.  The
supersymmetric vacua \susyvacuum\ are separated from the non-supersymmetric one at the
origin by a potential barrier. The vacuum at the origin is long-lived for $\mu\ll\Lambda$.

\subsec{Finite temperature}

The authors of \refs{\AbelCR\CraigKX\FischlerXH-\AbelMY} asked whether early universe
dynamics is likely to drive the ISS theory to the metastable state at $\Phi=0$, or to
the lower energy supersymmetric state \susyvacuum. They assumed that after inflation
and reheating the universe is in a thermal state with an adiabatically decreasing
temperature.\foot{For this to be true, the thermal equilibration time must be much
shorter than other relevant time scales, such as the rate of change of the temperature,
and the Hubble time.} In that case, one needs to evaluate the finite temperature corrections
to the effective potential $V_0+V_1$, \vveeff, \voneloop. The thermal effective potential $V_T$,
which depends non-trivially on both the pseudo-moduli $X$, \mmqq, and magnetic quarks, $q$,
$\tilde q$, takes the form
\eqn\potT{V_T = V_{0} + V_1+ V_1^{(T)},}
where \DolanQD
\eqn\vToneloop{V_1^{(T)}(q,\tilde q,X) = \frac{T^{4}}{2\pi^{2}}\sum_{i}\pm n_{i}\int_{0}^{\infty}d p\,
p^{2}\ln\left(1\mp\exp(-\sqrt{p^{2}+m_{i}^{2}(q,\tilde q,X)/T^{2}})\right)~.}
Here $n_i$ and $m_i$ label the degeneracies and masses of different fields, respectively. The sum
is over all the fields in the theory, with the upper sign for bosons and the lower one for fermions.

For temperatures slightly above
\eqn\thcritq{T_c \simeq {\mu\over\sqrt{N_c + 2 N_f}}~,}
the effective potential \potT\ has a global minimum at the origin of field space,
$q=\tilde q=\Phi=0$, and a local one at the location of the supersymmetric vacuum,
\susyvacuum. The global and local minima are separated by a potential barrier, just
like in the zero temperature case \IntriligatorDD. At these temperatures, thermal
fluctuations are not sufficient to push the system over the barrier. Let us assume for
now that at that point the system is in the basin of attraction of the global minimum at
the origin (we will revisit this assumption below), and consider its evolution.

When the temperature decreases below $T_c$, \thcritq, the origin becomes unstable
to condensation of the magnetic quarks $q$, $\tilde q$, and the system undergoes a
second order phase transition to a phase with broken gauge symmetry, which for low
temperatures approaches the metastable vacuum \mmqq. Initially, the supersymmetric
vacuum still has a higher free energy than the metastable one, but when the
temperature drops below a value of order $\sqrt{h}T_c$ (we are assuming that the
Yukawa coupling $h\ll 1$), the free energy of the supersymmetric vacuum becomes lower.

At that point, the metastable vacuum can decay  to the supersymmetric one by tunneling.
However, the time scale for this first order phase transition is governed by the parameter
$\mu/\Lambda$, which is taken to be small to ensure the metastability of the vacuum \mmqq\
at zero temperature. This ensures that the metastable vacuum does not decay in the early
universe as well, at least when the temperature is of order $T_c$ \thcritq\ or lower. Thus, at
late times, when the temperature $T\to 0$, the system remains in the metastable vacuum.

In the above discussion we assumed that at $T\sim T_c$ the system is near the metastable
vacuum at the origin. If, instead, it is near the supersymmetric vacuum, it would remain there at
late times. Thus, an important question is whether dynamics in the regime  $T>T_c$ leads
predominantly to one or the other outcome.

The authors of \AbelCR\ argued that high temperature dynamics drives the system to the
metastable vacuum.  The basic point is that as the temperature increases, the supersymmetric
vacuum becomes less stable, until it disappears at a temperature of order\foot{In \AbelMY\ it was
suggested that the temperature at which the SUSY vacuum disappears is somewhat lower than
$h\Phi_0$, but still much higher than $T_c$, \thcritq.} $h\Phi_0$, \susyvacuum.
If the reheating temperature after inflation is higher than this value, $\Phi$ rolls towards the
origin and is likely to get trapped there as the temperature decreases (as described above).

We see that the ISS system provides an example of dynamical vacuum selection. At zero
temperature it has two (types of) ground states, one of which is stable and supersymmetric
and the other metastable and non-supersymmetric. One might expect that the system will
end up at late times in the supersymmetric ground state, or assign some other probabilities
to the two vacua. Instead, one finds that dynamics favors the metastable vacuum over the
supersymmetric one. Below, we will generalize this picture to brane systems in string theory,
which, as mentioned above, have in some cases a more intricate vacuum structure.

\subsec{Generalized ISS model}

The unbroken R-symmetry of the ISS model leads to anomalously small gaugino masses in the
context of gauge mediation (see \eg\ \KomargodskiJF\ for a recent discussion). Thus, it is
interesting to consider R-symmetry breaking deformations of this model. An example of such
a deformation, studied in \refs{\GiveonEF,\EssigKZ}  (see also \refs{\GiveonYU,\KoschadeQU}), 
is obtained by replacing the magnetic superpotential \issw\ by
\eqn\wdef{W_m=h q\Phi\tilde q -h\mu^2\Tr\Phi+\half h^2\mu_\phi\Tr\Phi^2~,}
which gives rise to the bosonic potential
\eqn\vvdef{V_0=|h|^2\left(|q\tilde q-\mu^2I_{N_f}+h\mu_\phi\Phi|^2+|q\Phi|^2+|\Phi\tilde q|^2\right)~,}
We will restrict below to the case $\mu_\phi\ll\mu$, for which one can consider the last term in \wdef\
as a small perturbation of \issw.

To study the vacuum structure of the theory with deformed superpotential \wdef, it is convenient
to parameterize the light fields as follows:
\eqn\metaformphi{h \Phi=\left(\matrix{0_k & 0 & 0\cr 0 & h \Phi_n & 0 \cr 0 & 0 &
{\mu^2 \over \mu_\phi}I_{N_f-k-n}\cr}\right),\quad q\wt q   =\left(\matrix{\mu^2 I_k & 0 & 0\cr 0 &
\varphi\wt\varphi  & 0 \cr 0 & 0 &
0_{N_f-k-n}\cr}\right) ~.}
Here $\varphi$ and $\wt\varphi$ are $n\times (N_f-N_c-k)$ dimensional matrices, while  $\Phi_n$
and $\varphi\wt\varphi$ are  $n\times n$ matrices. This parameterization is motivated by the brane
construction, which will be reviewed in the next section.

Unlike the case $\mu_\phi=0$, the superpotential \wdef\ admits classical supersymmetric
vacua,\foot{For some values of $N_f$, $N_c$, there are  additional supersymmetric vacua
that are quantum in nature \GiveonEF.} which correspond in \metaformphi\ to $n=0$,
$k=0,1,\cdots, N_f-N_c$. In these vacua, the $SU(N_f-N_c)$ magnetic gauge symmetry is
broken to $SU(N_f-N_c-k)$, by the expectation value of the magnetic quarks. Some components
of $\Phi$ get a large $(\gg\mu)$ expectation value. The fields $\Phi_n$, $\varphi$,
$\wt\varphi$ are absent since $n=0$.

There are also metastable vacua, in sectors with $n\not=0$. In studying them, it is useful to
discuss separately two cases:
\item{(1)} For $k=N_f-N_c$ the magnetic gauge group is completely broken. The fields $\varphi$,
$\wt\varphi$ are absent in this case, so the $n\times n$ matrix $\varphi\wt\varphi$ in \metaformphi\
should be set to zero. The potential for $\Phi_n$ has two contributions. One is the classical
potential \vvdef, which pushes it towards the supersymmetric vacuum $h\Phi_n=(\mu^2/\mu_\phi)I_n$.
The other comes from the one loop potential \voneloop, and provides an attractive force towards the origin.
Balancing the two against each other, one finds \GiveonEF\ a local minimum of the potential near the origin, at
\eqn\metast{h\Phi_n\simeq {\mu_\phi\over N_f-N_c}~.}
In order for this vacuum  to be in the regime of validity of the approximation \voneloop, $\mu_\phi$
must be very small, $\mu_\phi\ll h\mu(N_f-N_c)$.
\item{(2)} For $k<N_f-N_c$, part of the $SU(N_f-N_c)$ gauge group is unbroken by \metaformphi, and the
quark fields $\varphi$, $\wt\varphi$ need to be included in the discussion. Looking back at \vvdef\
one finds that they are tachyonic, unless the minimum of the potential for the pseudo-moduli \metast\
is at $\Phi_n>\mu$. In this regime, the approximation to the one loop potential \voneloop\ is not valid.
The full one loop potential has no metastable vacua in these sectors \EssigKZ\ (see also \GiveonYU).

\noindent
As shown in \GiveonEW, the brane system that realizes the generalized ISS field theory in a certain
regime in its parameter space has a similar phase structure, but it has additional metastable states
in sectors with $k<N_f-N_c$. In the next section we review this construction and its vacuum structure.

The discussion of \refs{\AbelCR\CraigKX\FischlerXH-\AbelMY} of finite temperature dynamics of
the ISS model is easy to generalize to the case of non-zero $\mu_\phi$. The vicinity of the metastable
vacuum \metast\ can be studied as in the previous subsection, since the small $\mu_\phi$ does not
modify the potential by much (see \GiveonBV\ for a related discussion). The fate of the
supersymmetric vacua discussed above, \metaformphi\ with $n=0$, can also be studied.
The situation here is better than in  \AbelCR, since these vacua are visible classically in the WZ
model, and do not require non-perturbative effects in the non-asymptotically free magnetic theory.
By evaluating the effective potential \vToneloop, one finds that the SUSY vacuum disappears
when the temperature exceeds a value of order $\mu^2/\mu_\phi$ (with a small coefficient, of
order $0.01-0.1$ for the values of the parameters that we considered).

\newsec{Review of the brane construction}

In this section we review the brane construction of the ISS model
\refs{\OoguriBG\FrancoHT\BenaRG-\GiveonFK}, and its generalization presented
in \GiveonEW. To describe it, it is convenient to decompose the $9+1$ dimensional
spacetime of type IIA string theory as follows:
\eqn\decomp{\IR^{9,1}=\IR^{3,1}\times \IC_v\times \IC_w\times\IR_y\times\IR_{x^7}~,}
with
\eqn\nsns{v=x^4+ix^5~,\qquad w=x^8+ix^9~,\qquad y=x^6~.}
The brane configuration we consider is depicted in figure 1. All branes are extended in the
$\IR^{3,1}$ labeled by $(x^0, x^1,x^2,x^3)$. The $NS5$-branes denoted by $NS$ and $NS'$ are
further extended in $v$ and $w$, respectively. The  $D4$-branes are stretched between other
branes as indicated in the figure.

\ifig\loc{The ISS brane configuration.}
{\epsfxsize5.0in\epsfbox{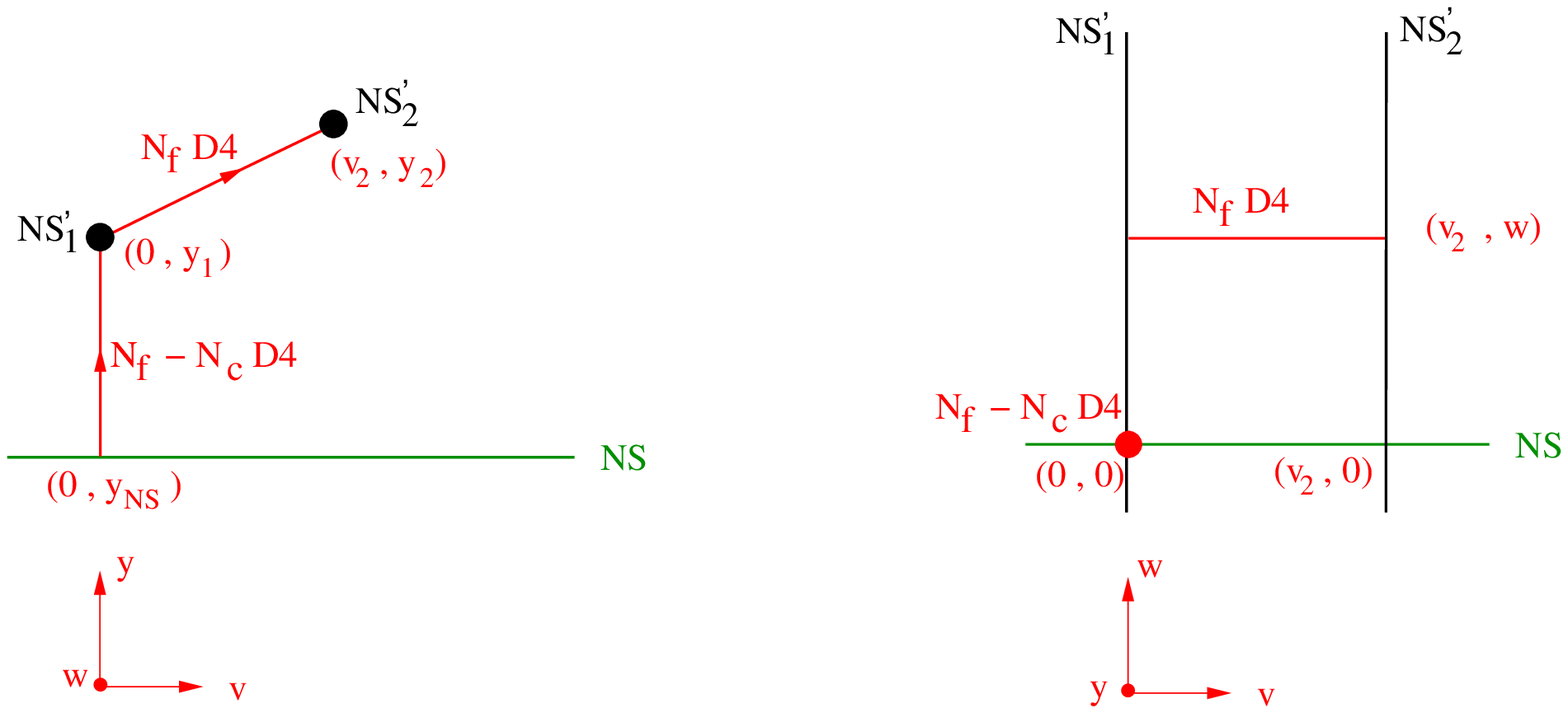}}

\noindent
The fivebranes alone preserve $N=2$ supersymmetry in four dimensions. They can be
thought of as providing a dual description of a non-compact Calabi-Yau (CY) geometry.
For $y_i=y_{NS}=v_2=0$ this geometry is singular, and can be described as in \GiveonZM.
Varying these parameters corresponds to smoothing out the singularity. The $D4$-branes
stretched between $NS5$-branes in figure 1 correspond in the dual picture to $D5$-branes
wrapping small two-cycles in the non-compact CY. The CY description is useful when one
of the directions transverse to the fivebranes is compactified on a small circle. We
will consider the opposite limit, in which the fivebrane picture is the appropriate one.

At low energies, the brane configuration of figure 1 reduces to an $N=1$ supersymmetric
gauge theory with gauge group $SU(N_f-N_c)\times SU(N_f)\times U(1)$, $N_f$ flavors of
bifundamentals $q,\tilde{q}$ with opposite $U(1)$ charges, and an adjoint of $SU(N_f)$,
$\Phi$. In the brane picture, the eigenvalues of $\Phi$ correspond to the locations in
the $w$-plane of the $D4$-branes stretched between $NS'$-branes. The precise relation
between $\Phi$ and $w$ is described \eg\ in \GiveonUR.

The chiral superfields $q$, $\tilde q$ and $\Phi$ have canonical K\"ahler potential,
and superpotential
\eqn\wmaggg{W_{\rm mag}=h  q\Phi \tilde q- h \mu^2 \Tr\Phi~.}
This theory is essentially the same as that described in section 2.1, except for the
gauging of the flavor symmetry group. The theory in which the flavor group is not
gauged can also be realized as a low energy theory on branes, by replacing the
$NS'_2$-brane in figure 1 by $N_f$ $D6$-branes \ElitzurFH. For our purposes
below it will be convenient to consider the system with gauged flavor group.

The parameters in the low energy theory on the branes can be expressed in terms of
string theory quantities as follows. The gauge couplings $g_{YM}^2$ for the different
factors of the gauge group are proportional to $g_s$ divided by the extent in $y$ of the
relevant fourbranes \GiveonSR. The parameters in the superpotential \wmaggg\ are given by
\eqn\hmug{h^2={8\pi^2g_sl_s\over \Delta y}~,\qquad \mu^2={v_2\over 16\pi^3 g_s l_s^3}~,}
where $\Delta y=y_2-y_1$, and we made a choice of phase for the chiral superfields such
that the complex couplings $h$ and $\mu$ are real and positive.

The geometric description is reliable when the distances between the various branes are
large (relative to $l_s$), and the string coupling $g_s$ is small. In this regime, the
gauge couplings and Yukawa coupling $h$ are small. The mass $\mu$, which is
determined by the geometric parameter $v_2$, is typically above the string scale. If
$\mu$ is sufficiently small, the low energy dynamics of the branes is well described
by magnetic SQCD. In general, the low energy effective Lagrangian receives contributions
from other sources \refs{\GiveonFK,\GiveonUR}.

\ifig\loc{The marginally stable brane configuration.}
{\epsfxsize5.0in\epsfbox{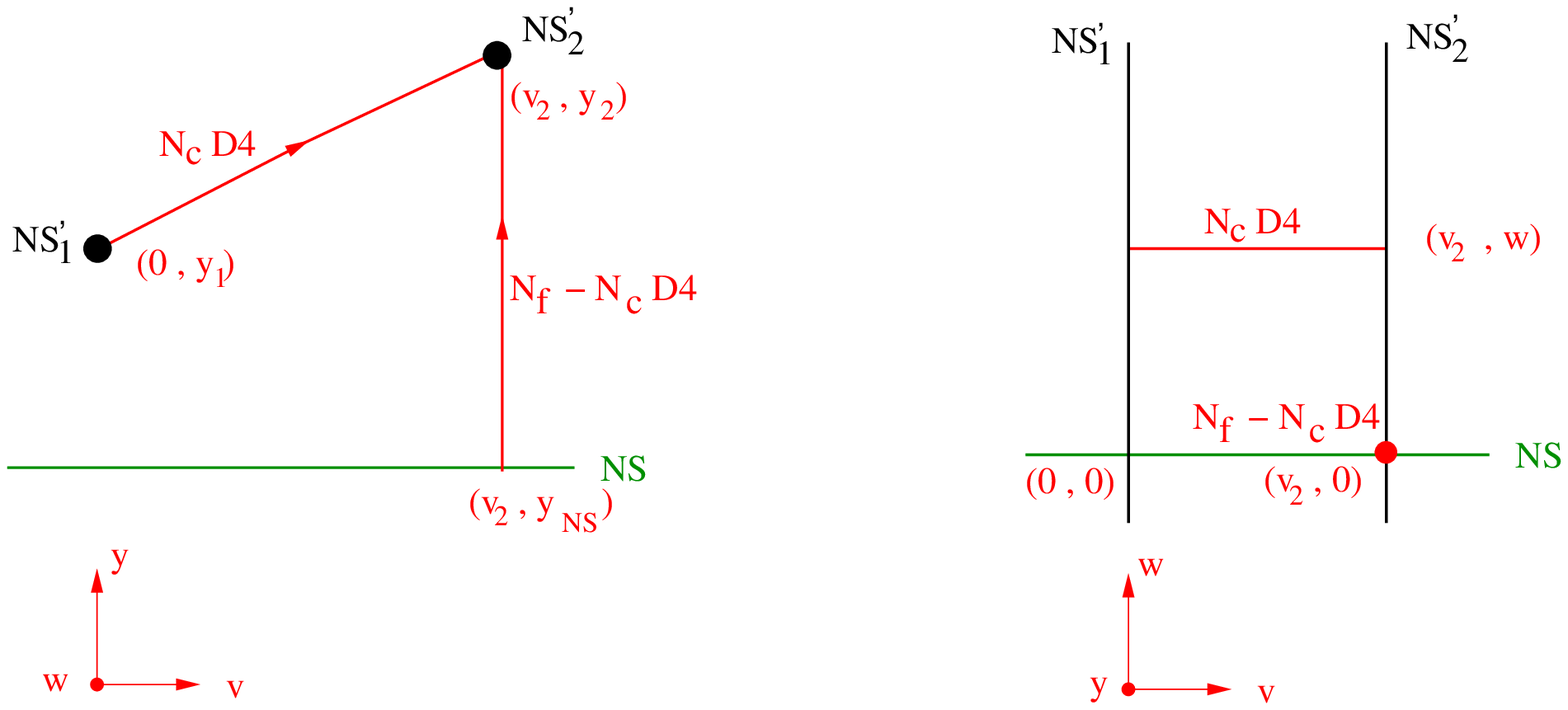}}

The brane configuration of figure 1 is unstable to reconnection of the  $N_f-N_c$ ``color''
$D4$-branes with some of the $N_f$ ``flavor'' $D4$-branes, leading to that of figure 2.
The resulting configuration is marginally stable. It contains an $N_c\times N_c$ matrix
$X$ of massless fields with a flat potential. The (pseudo-) moduli space labeled by $X$ is
a direct analog of that found in the low energy field theory analysis in section (2.1).
As discussed after eq. \mmqq, the exact flatness of the potential for $X$ may be spoiled
by small corrections to the dynamics.

In the field theory regime (\ie\ when the geometric parameter $v_2$ in figure 2 is small,
so that  $\mu\ll m_s$), the leading correction is the CW potential \voneloop, which lifts
the moduli space by giving a mass to $X$. In string theory language, the CW potential
\voneloop\ is a string one loop effect, and is thus suppressed in the $g_s\to 0$ limit.
In the regime where all distances in figure 2 are large in string units, the dominant
contribution to the potential for $X$ comes from the classical gravitational attraction of
the $N_c$ $D4$-branes to the $NS$-brane \GiveonFK. From the point of view of the low energy
effective field theory for $X$, this attraction gives rise to non-renormalizable D-terms
\GiveonUR\ that, together with the linear superpotential \issw, provide a potential which
stabilizes $X$ at the origin.

Thus, the fate of the pseudo-moduli space labeled by $X$ is the same in the field theory
and string theory regimes -- it is replaced by an isolated supersymmetry breaking vacuum
at the origin -- but the mechanisms are different. The supersymmetric
vacuum \susyvacuum\ is not visible in the classical brane description. In string theory it
owes its existence to quantum ($g_s$) effects, and is thus located at a very large value of $w$.

So far in this section we reviewed the realization of the ISS model as a low energy theory
on branes. Since our main interest is in the generalized ISS model described in section 2.3,
we next turn to the corresponding  brane system. The brane configuration, which is depicted
in figure 3, is obtained from that of figure 2 by a rotation of the $NS'_2$-brane by an angle
$\theta$ in $(v,w)$. The parameters $h$, $\mu$ are again given by \hmug, while
\eqn\muphi{ \mu_\phi = {\tan{\theta} \over 8\pi^2 g_s l_s }~.}

\ifig\loc{The deformed ISS brane configuration.}
{\epsfxsize5.0In\epsfbox{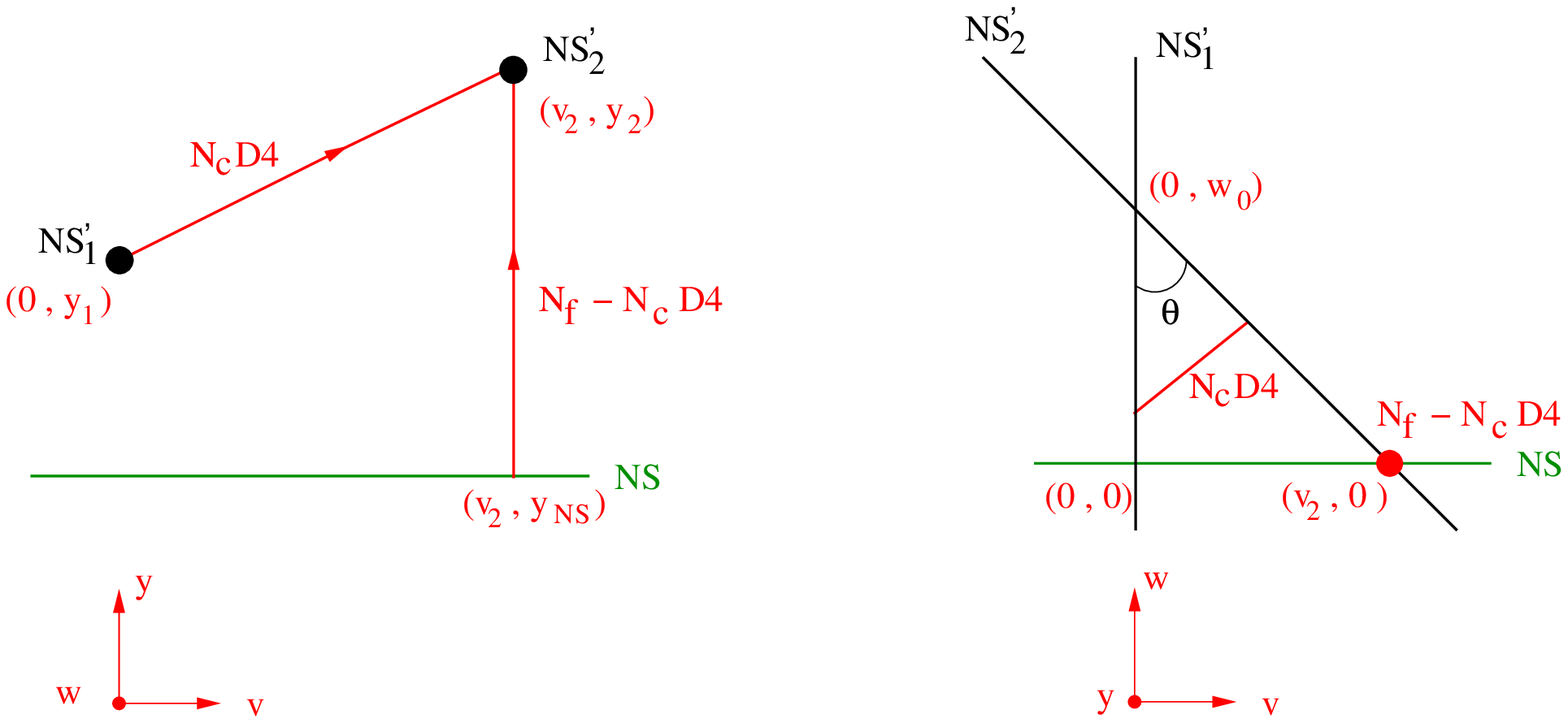}}

\noindent
The orientation of the $NS'_2$-brane leads to an important difference
in the dynamics of the configuration of figure 3 compared to that of
figure 2. While the latter has (to leading order) a pseudo-moduli
space of supersymmetry breaking vacua labeled by the positions of the
$N_c$ $D4$-branes in $w$, in the former this moduli space is lifted
and replaced by a supersymmetric vacuum, in which the $D4$-branes are
located at the point $(v,w)=(0,w_0)$, with
\eqn\intbranes{w_0=v_2\cot\theta~.}
At that point, the projections of the $NS'$-branes on the $(v,w)$
plane intersect, and the $D4$-branes stretched between them preserve
$N=1$ supersymmetry.

In fact, the brane system of figure 3 has many supersymmetric vacua, which
were described and compared to field theory in \GiveonEW. From the perspective
of the WZ model with superpotential \wdef, these vacua are obtained by analyzing
the zeroes of the bosonic potential \vvdef. The brane system also has many
non-supersymmetric vacua \GiveonEW.

The classical supersymmetric and non-supersymmetric  vacua are exhibited in
figure 4. They are labeled by two integers  $k$, $n$, which are the same as
those defined in the field theory in eq. \metaformphi. The $N_f-N_c-k$
$D4$-branes stretched between the $NS$ and $NS'_1$-branes give rise to an
unbroken $SU(N_f-N_c-k)$ subgroup of the magnetic gauge group. A
$SU(k)\times SU(n)\times SU(N_f-k-n)$ subgroup of the gauge group is unbroken
as well.

\ifig\loc{The vacuum structure of the deformed brane configuration.}
{\epsfxsize5.0in\epsfbox{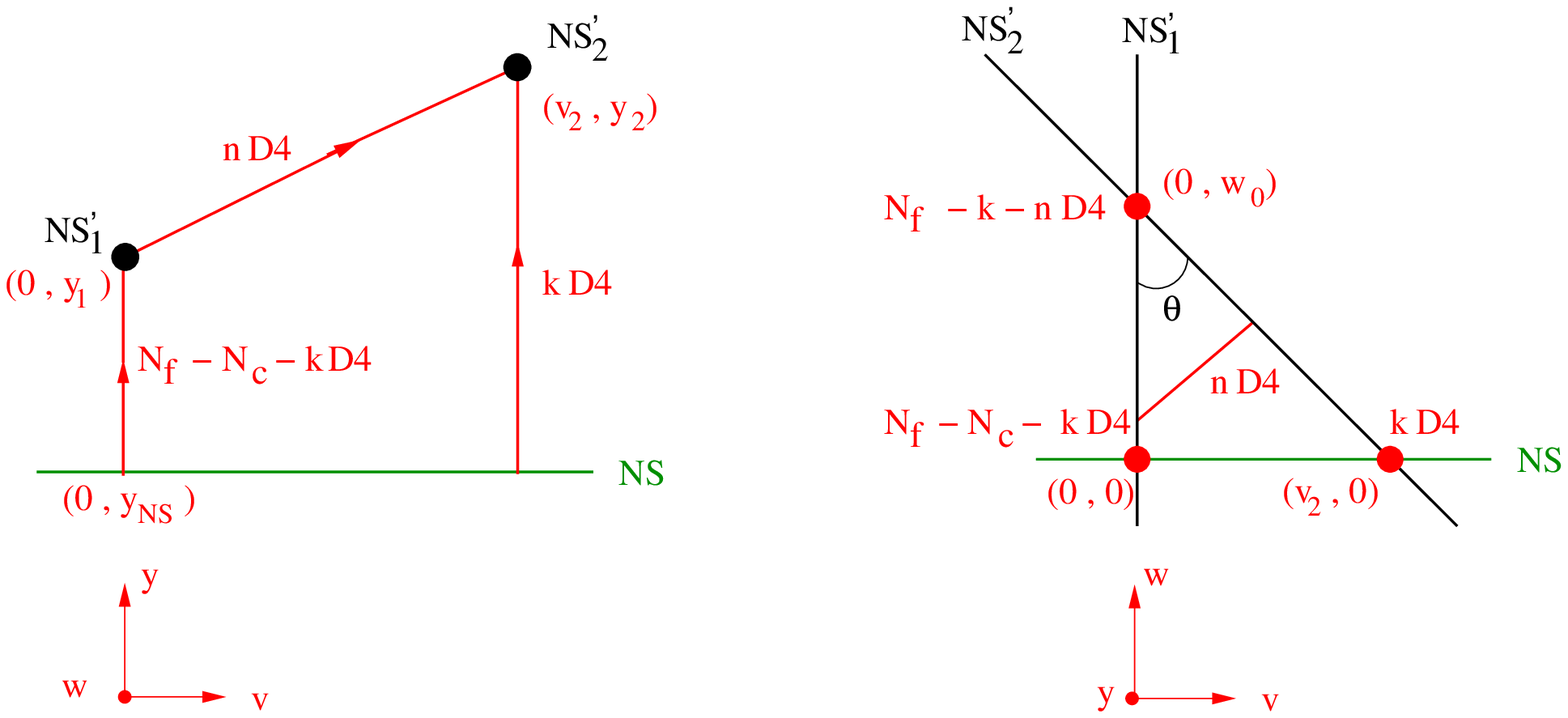}}

In the low energy field theory, these vacua can be understood in terms of the
decomposition \metaformphi. As mentioned there, the supersymmetric vacua
correspond to $n=0$. They can be obtained by moving the $n$ $D4$-branes
in figure 4 to the point $(v,w)=(0,w_0)$, where they join the $N_f-k-n$ $D4$-branes
already there.

The vacua with $n>0$ seem at first sight unstable, since the $n$ $D4$-branes
can decrease their energy by moving towards the aforementioned supersymmetric
vacua. However, the attraction of the $D4$-branes towards the $NS$-brane
provides a counterbalancing force. The combination of the two effects leads to
the existence of a local minimum of the energy at a value of $w$ which was
computed in \GiveonEW, and interpreted from the low energy field theory point
of view in \GiveonUR. These non-supersymmetric vacua can be made parametrically
long-lived by tuning $g_s$ and other parameters of the brane configuration.

As is clear from figure 4, there are two qualitatively different types of metastable
vacua with $n>0$. For vacua with $k=N_f-N_c$, the only potential mode of instability
is the tunneling of the $n$ fourbranes through the potential barrier in the $\Phi$
direction. The rate for this process can be suppressed by tuning $g_s$ and other
parameters of the brane configuration.

For $k<N_f-N_c$, the brane configuration of figure 4 has another mode of instability.
Some or all of the $n$ $D4$-branes can connect with color $D4$-branes stretched
between the $NS$ and $NS'_1$-branes and move towards the $k$ $D4$-branes stretched between
the $NS$ and $NS'_2$-branes. Depending on whether $n$ is larger or smaller than $N_f-N_c-k$,
this process leads to metastable vacua of the previous kind, or to supersymmetric vacua
with a different value of $k$.

The reconnection process described above is a tachyonic instability or a tunneling
process, depending on the distance between the ends of the various $D4$-branes
on the $NS'_1$-brane in figure 4. That distance is in turn a function of the parameters
of the brane configuration. It was shown in \GiveonEW\ that in the geometric brane
regime one can choose the parameters such that there is no tachyonic instability.
In this regime, all vacua in figure 4 are locally stable. Of course, the vacua with
$k<N_f-N_c$ have shorter lifetimes than those with $k=N_f-N_c$, due to the
additional decay modes. All lifetimes can be made arbitrarily large by decreasing
the string coupling $g_s$.

From the effective field theory point of view, the difference between the cases $k=N_f-N_c$
and $k<N_f-N_c$ is the absence or presence of the fields $\varphi$, $\wt\varphi$ in the
decomposition \metaformphi. These fields correspond in figure 4 to strings stretched between
the $n$ and $N_f-N_c-k$ $D4$-branes. Their potential includes a tachyonic contribution at tree
level, and a positive one due to the expectation value of the field $\Phi_n$ at the local
minimum of its potential. In the brane regime, the potential for $\Phi_n$ is due to classical
gravitational interactions, which are strong enough to push the system out of the tachyonic
region \refs{\GiveonEW,\GiveonUR}. In the field theory regime, where this potential is due to
one loop effects in the low energy WZ model, the vacua with $k<N_f-N_c$ are always
destabilized by $\varphi$ condensation \EssigKZ.

\newsec{Early universe dynamics}

At early times, the brane systems described above are expected to be in an excited
state. The question we would like to address is what happens to such a state as it
evolves and, in particular, which of the vacua described in section 3 is approached by the system
at late times. In this section we will address this question, first for
the ISS system of figures 1, 2, and then for the generalized ISS system of figures 3, 4.

In a typical excited state of a brane system of the sort described
in section 3, the excess energy can go into exciting the fivebranes
and/or the $D$-branes. It is natural to expect that for small string coupling,
most of the energy goes into exciting the fivebranes, since they are the
heaviest objects around. A generic excited state of an $NS5$-brane
corresponds to a charged black brane. Thus, we will discuss the brane
systems of section 3 with the $NS$ and $NS'$-branes taken to be 
non-extremal.\foot{DK thanks S. J. Rey for describing his unpublished work
\sjrey\ on Hanany-Witten constructions with non-extremal fivebranes, in June 2008.}

The non-extremality of the fivebranes leads to modifications of the potential
energy landscape. Our main purpose below is to map out this  landscape.
Once this is understood, one needs to study the real time dynamics of the
$D4$-branes in this landscape and, in particular, analyze the possible late
time limits of the time evolution. We will briefly comment on this below, leaving
a more complete study to future work.

In the analysis \refs{\GiveonFK,\GiveonEW} of the brane systems of section 3,
a dominant role was played by the gravitational attraction of the flavor $D4$-branes
to the $NS$-brane. We will start by analyzing the modification of this attraction
due to the non-extremality of the $NS$-brane, and then take into account some
additional effects.

The background of an excited $NS$-brane, which wraps the $\IR^{3,1}$ labeled
by $(x^i,t)$ and the $v$-plane, is given by the non-extremal fivebrane
solution\foot{There is also a non-zero $NS$ $H$ field proportional to the volume
form on the transverse $S^3$.} \HorowitzCD:
\eqn\nonex{ \eqalign{ & ds^2 = - f(r) dt^2 + H(r) [ f(r)^{-1} dr^2 + r^2 d\Omega_{3}^2 ] +
(dx^i)^2 + dv d \vbar~, \cr
& e^{2(\varphi-\varphi_0)} = H(r) ~.}}
Here $(r, \Omega_3)$ are spherical coordinates on the $\IR^4$ transverse to the $NS$-brane.
The radial coordinate is given by $r^2 = (y-y_{NS})^2 + (x^7)^2 + w \wbar$.
The functions $f,H$ are given by
\eqn\fH{ f(r) = 1 - \frac{r_h^2}{r^2} ~, ~~~~~ H(r) = 1 + \frac{l_{s}^2}{r^2}~.}
$\varphi$ is the dilaton field; $e^{\varphi_0}=g_s$ is the string coupling far from the
fivebranes.

The geometry \nonex\ describes a black brane with a horizon at $r = r_h$.
The size of the black brane $(r_h)$ increases with its energy density, $\EE$,
as follows
\eqn\enden{ \EE = \frac{1}{(2\pi)^5 l_{s}^6 g_{s}^2} \left( 1 + \frac{r_h^2}{l_{s}^2} \right)~.}
As $r_h \rightarrow 0$, \nonex\ approaches the extremal fivebrane (CHS) geometry
\CallanAT, and \enden\ approaches the tension of a BPS $NS5$-brane. For $r_h\gg l_s$,
the energy of the non-extremal fivebrane is much larger than its charge, and one can set
$H(r)\simeq 1$. This leads to the $4+1$ dimensional Schwarzschild black hole (extended
in the five dimensions $(x^i,v)$). Its Hawking temperature is given as usual by $T_h=1/r_h$.

The non-extremal brane loses energy via Hawking radiation. As a result, its horizon shrinks
at the rate \MaldacenaCG
\eqn\Hawking{ {d\over dt}(r_h^2) \sim - g_s^2l_{s} ~.}
If the initial radius is of order $l_s$ or larger, eq. \Hawking\ implies that it will take a time
$t \sim l_s/g_{s}^2$ or more  to lose its excess energy. This is natural, since brane decay
is a quantum process, and is thus suppressed by a closed string loop factor. The decay
rate of a $D$-brane is proportional to $g_o^2\sim g_s$. The decay time of a $D$-brane is thus
$\sim l_s/g_s$, which is much shorter than that for an $NS5$-brane (for $g_s\ll 1$).

The dynamics of the $D$-branes in figure 2 in the background of the extremal fivebrane
(\nonex\ with $r_h=0$) was analyzed in \GiveonFK. Our purpose here is to generalize this
analysis to $r_h\not=0$. We will split the discussion into two parts. In the next
subsection we discuss the case $\mu=0$ (see \issw); in the following one we move on to
the system of figure 2, in which $\mu\not=0$.

\subsec{$\mu = 0$}

As discussed in section 3, the system with $\mu=0$ in \issw\ corresponds in the brane
language to the configuration of figure 1, with the displacement of the $NS'_2$-brane
$v_2$ set to zero. This is the magnetic SQCD brane configuration of \ElitzurFH\ (with
gauged flavor group \refs{\GiveonSR,\GiveonFK}). It has a moduli space of supersymmetric
vacua labeled by the expectation value of $\Phi$ \issw. In the brane construction, this moduli
space corresponds to displacements in the $w$-plane of the $N_f$ flavor $D4$-branes \GiveonSR.

Making the $NS$-brane non-extremal (\ie\ turning on a non-zero $r_h$ \fH) breaks super-symmetry
and generates a potential on  moduli space. In order to compute it to leading order in $g_s$,
it is sufficient to consider a single $D4$-brane stretched in $y$ between the $NS'$-branes,
and calculate its energy as a function of $w$ in the background \nonex.

The Dirac-Born-Infeld (DBI) action for a $D4$-brane in the background of an $NS$-brane is
given by
\eqn\DBI{ \SS = -T_4 \int d^4 x \int_{y_1}^{y_2}dy e^{-\varphi} \sqrt{-{\rm det}P(G)_{ab}}~,}
where $T_4/g_s=1/(2\pi)^4l_s^5g_s$ is the tension of the $D4$-brane, and $P(G)$ is the pullback
of the spacetime metric \nonex\ to the worldvolume of the brane.  The potential is determined by
taking $w(x^\mu)$ to be constant,
\eqn\noaPot{ V(w) =  \frac{T_4}{g_s} \int_{y_1}^{y_2} dy
\sqrt{ 1 - \frac{ r_h^2 |w|^2 }{( (y - y_{NS})^2 + |w|^2 )^2 } } ~.}
As in \GiveonUR, we take the limit $y_i, \Delta y\ll|y_{NS}|$, in which the $D4$-brane
becomes a local probe of the geometry, and \noaPot\ simplifies to
\eqn\noaPottwo{ V  \simeq \frac{T_4 \Delta y}{g_s} \sqrt{ 1 - \frac{ r_h^2 |w|^2 }
{ (y_{NS}^2 + |w|^2 )^2 } }\simeq\frac{T_4 \Delta y}{g_s}
\left[1 - \half\frac{ r_h^2 |w|^2 }
{ (y_{NS}^2 + |w|^2 )^2 }\right]  ~.}
In the last step we expanded the potential to leading order in $r_h/y_{NS}$. We will
restrict the discussion to the region $l_s\ll r_h\ll \Delta y\ll |y_{NS}|$, and work to leading
order in the corresponding small parameters. This allows us to neglect some subleading
effects. For example, we have taken the $w$-position of the $D4$-brane to be independent
of $y$, \ie\ assumed that the $D4$-brane continues to stretch in $y$ when $r_h > 0$.
One can show that corrections to this approximation are down by a power of $r_h/y_{NS}$.

So far we assumed that the length of the $D4$-brane, $\Delta y$, is independent of $w$.
This is certainly the case for $r_h=0$, in which the $NS'$-branes are located at a fixed
value of $y$. However, in the non-extremal case they are deformed by the gravitational
potential of the $NS$-brane. To complete the calculation of the potential \noaPottwo\
we need to take this deformation into account.

The shape of an $NS'$-brane in the $NS$ background \nonex\ is determined by a Nambu-Goto
type action. The equations of motion of this action, which are  analyzed in appendix A, lead to the
shape
\eqn\NSsmallw{ y =y_i + (y_i-y_{NS})\frac{ r_{h}^2}{4 ((y_i-y_{NS})^2+ |w|^2)}~,}
where $y_i$ gives the asymptotic position of the $NS_{i}'$-brane with $i=1,2$ in figures 1--4.
The shape of the $NS'$-branes in the background of the $NS$-brane makes the
length of a $D4$-brane stretched between the two $NS'$-branes $w$ dependent.
Using \NSsmallw\ we find (to leading order in $y_i/y_{NS}$)
\eqn\Delysmallw{ \Delta y(|w|) \simeq \Delta y \left[ 1 +{r_h^2\over 4(y_{NS}^2+|w|^2)}
-{r_h^2y_{NS}^2\over 2(y_{NS}^2+|w|^2)^2}\right]~.}
Plugging this into \noaPottwo, we find the potential
\eqn\fullpotsmall{V(w)\simeq {T_4\Delta y\over g_s}\left[1-{r_h^2\over 4(y_{NS}^2+|w|^2)}\right]
\simeq{T_4\Delta y\over g_s}\left[1+{1\over4}{r_h^2\over y_{NS}^4}|w|^2
+\OO(|w|^4)\right]~.}
We see that the net effect of the non-extremality is to stabilize the $D4$-brane at the origin.

Some comments on the above analysis are in order:
\item{(1)} We focused on a single $D4$-brane stretched between $NS'$-branes in the non-extremal
fivebrane background \nonex. Coming back to the brane configuration of figure 1 with $v_2=0$,
the above analysis implies that for finite $r_h$ the $N_f$ flavor branes are localized at
$w=0$. The supersymmetric moduli space is replaced by a vacuum with unbroken gauge symmetry,
like in the corresponding field theory analysis.
\item{(2)}  We took the $NS'$-branes, between which the $D4$-brane stretches, to be extremal.
As mentioned above, it is more natural to take them to be non-extremal as well, with the
same horizon size as the $NS$-brane. A $D4$-brane stretching between two non-extremal
$NS'$-branes with horizon sizes $r_h$ can be thought of as ending on the horizons. Thus,
its length is smaller by $2r_h$ than in the extremal case. This does not change the
above discussion, since the resulting shift in energy is independent of $w$.
\item{(3)} Although the non-extremality of the $NS'$-branes does not qualitatively affect the
$D4$-branes, it nevertheless has an important effect on the dynamics of the system. For finite
$r_h$, the two parallel $NS'$-branes attract each other and merge after a time that scales like
$(g_s)^0$, much shorter than the scale of variation of $r_h$  (see \Hawking). Thus, the fivebrane
configuration of figures 1,2 is a very unnatural endpoint of the time evolution of this
system.\foot{This conclusion can be avoided either if the $NS'$-branes are initially at a very
large distance, so that their motion towards each other takes longer than the time during
which $r_h$ approaches zero, or if the rate of decrease of $r_h$ is in fact faster than that
due to Hawking radiation.}

\subsec{ISS system}

We now turn to the case  $\mu\not=0$, which is described in terms of branes by the
configuration of figures 1, 2. Recall first the situation in the extremal case $r_h=0$. The brane
system of figure 1 appears to have flat directions corresponding to the locations of the
$N_f$ flavor $D4$-branes in the $w$-plane. However, gravitational attraction to the $NS$-brane
provides a potential for these fourbranes, which localizes them at the origin \GiveonFK.

At that point the magnetic quarks, corresponding to strings connecting color and flavor branes,
are tachyonic. Condensation of these tachyons corresponds to a reconnection process,
which leads to the configuration of figure 2. The $N_c$ flavor branes that remain between
the $NS'$-branes are still localized at the origin, due to the gravitational potential of
the $NS$-brane. Thus, the resulting non-supersymmetric vacuum is locally stable.

We would like to extend the above picture to the non-extremal case $r_h\not=0$. Consider first
the potential on pseudo-moduli space. Following \GiveonUR\ we take the limit $y_i,\Delta y\ll|y_{NS}|$,
in which the shape of the $D4$-brane in the $(v,y)$ plane is well approximated by a straight line
\eqn\adef{ v = a (y - y_1)~.}
We assume that the slope parameter,
\eqn\smallaa{a = \frac{v_2}{\Delta y}~,}
is small. The potential $V(w)$ obtained from the DBI action \DBI\ is given to leading order in
$r_{h}/y_{NS}$ and $a$ by
\eqn\nothetaPot{ \eqalign{ V(w) = &   \frac{T_4}{g_s} \int_{y_1}^{y_2} dy \left[ 1 - \frac{ r_h^2 |w|^2 }
{( (y - y_{NS})^2 + |w|^2 )^2 } + a^2 \frac{ (y - y_{NS})^2 + |w|^2 - r_h^2 }
{ ( y - y_{NS})^2 + |w|^2 + l_{s}^2 } \right]^{1/2}  \cr
\simeq &   \frac{T_4}{g_s}  \Delta y(|w|)
\left[ 1 -\half \frac{ r_h^2 |w|^2 }{( y_{NS}^2 + |w|^2 )^2 } +
{a^2\over2} \frac{ y_{NS}^2 + |w|^2 - r_h^2 }
{ y_{NS}^2 + |w|^2 + l_{s}^2 } \right] ~.}}
Taking into account the bending of the $NS'$-branes, \Delysmallw, we find for $|w| \ll |y_{NS}|$:
\eqn\nothetaPottwo{ V(w) \simeq \frac{T_4 \Delta y}{g_s} \left[ 1 +
\frac{ (r_{h}^2 + 2 a^2 l_{s}^2) }{4 y_{NS}^4} |w|^2+ \OO(|w|^4)  \right]~.}
We see that, as one would expect, near the origin of pseudo-moduli space the potential is a sum
of the attractive contributions due to the two sources of supersymmetry breaking -- non-zero $r_h$
\fH, and non-zero $a$, \adef. For $r_h\gg l_s$ the contribution from non-extremality is much larger
than the other one.

The second element of the discussion of the extremal case that needs to be reexamined for
$r_h\not=0$ is the reconnection of color and flavor branes, once the latter are localized
at $w=0$. If the $NS'_1$-brane is extremal, the situation is as before; $N_f-N_c$ flavor
branes connect to the color branes and slide off towards the configuration of figure 2.
However, as mentioned above, early universe dynamics is expected to give rise to non-extremality
of the $NS'$-branes as well. Their horizon radii are likely to be comparable to that of the
$NS$-brane; for simplicity we will take them to be equal below.

Thus, we need to take the $NS'_1$-brane to be non-extremal with horizon radius $r_h$, and
repeat the analysis of the tachyonic instability associated with the reconnection process
mentioned above. This problem is studied in appendix B; here we summarize the results.

\ifig\loc{Stationary points of the DBI action for a $D4$-brane for  different value of $r_h$. The 
short-dashed blue line corresponds to a gauge symmetry breaking local minimum, the solid 
red line is a saddle point, and the long-dashed green line a symmetry preserving local minimum. 
The parameters in the text were chosen as follows: $a = 0.1$, $\Delta y = 0.1 |y_{NS}|$, 
$r_h/v_2=0.2, 0.136, 0.025$ for figures (a),(b),(c), respectively.}
{\epsfxsize5in\epsfbox{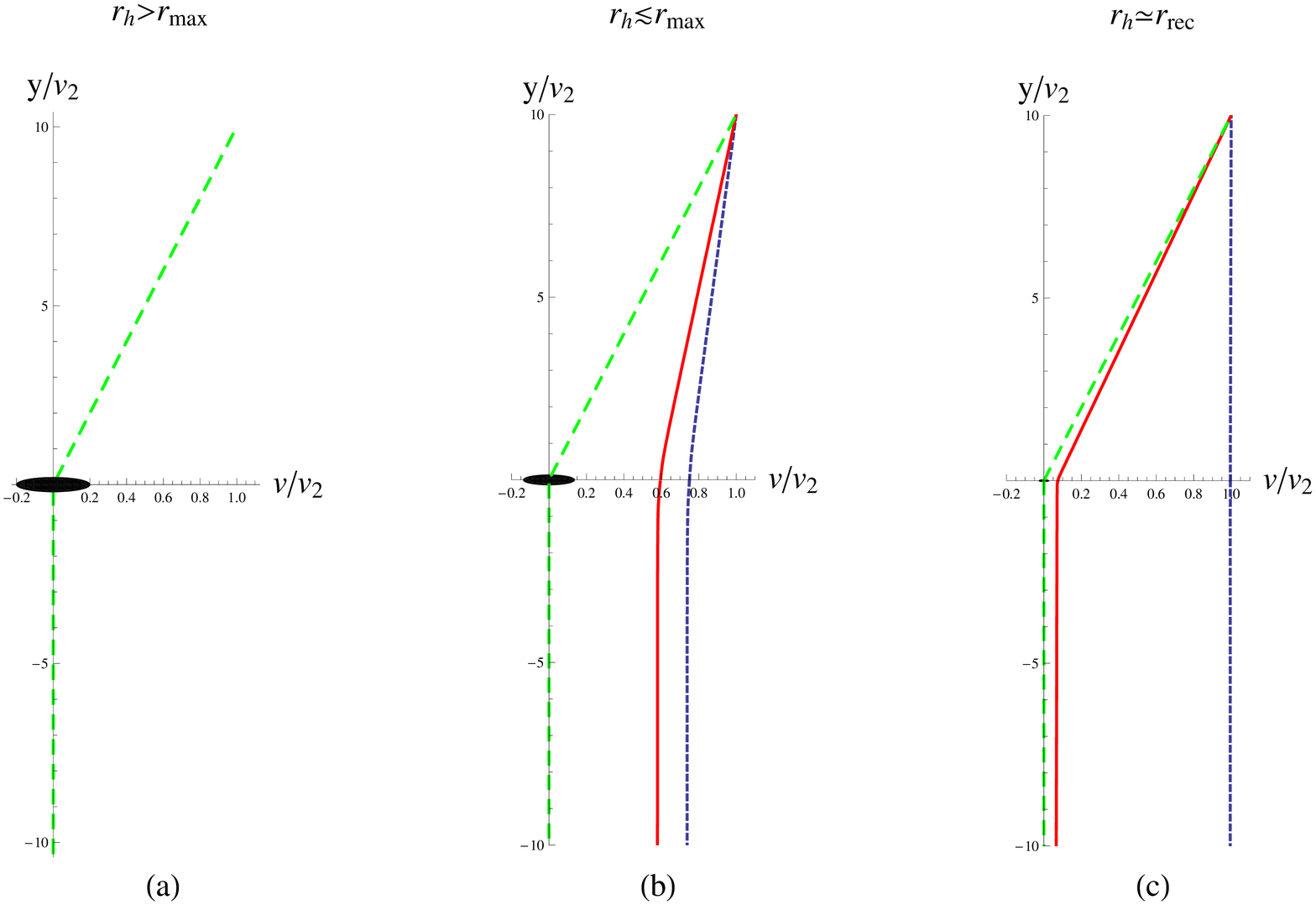}}

For
\eqn\RmaxText{ r_h > r_{\rm max} \simeq  \frac{4 a^{3/2}\Delta y }{3 \sqrt{3\pi}}  =
\frac{4 }{3 \sqrt{3\pi}} \sqrt{a} v_2  ~,}
the configuration denoted by a long-dashed green line in figure 5, which consists of
disconnected color and flavor branes ending on the horizon of the $NS_1'$-brane,
is the only solution of the equations of motion for the $D4$-brane. In particular, in this 
range of $r_h$, the brane reconnection instability discussed above is absent, and the 
gauge symmetry is unbroken.

For $r_h<r_{\rm max}$, two additional stationary points of the DBI action for the
$D4$-brane appear. These configurations, denoted by the solid red and short-dashed
blue lines in figure 5, correspond to a saddle point and local minimum of the energy,
respectively. They break part of the gauge symmetry of the brane system. The saddle
point (solid red line in figure 5) has, by construction, a higher energy than the gauge
symmetry preserving configuration (long-dashed green line), and local minimum
(short-dashed blue line). When $r_h$ is in the range
\eqn\RrecText{
r_{\rm max}>r_h>r_{\rm rec} \simeq
\frac{a^2}{4} \Delta y = \frac{a}{4} v_2~,}
the symmetric configuration has lower energy than the symmetry breaking one. For
$r_h<r_{\rm rec}$, the minimum energy configuration is the symmetry breaking one.
The energetics of this brane system is summarized in figure 6.

\ifig\loc{Energies of the brane configurations of figure 5 as a function of $r_h$
(same parameters and color/dashing scheme as there).}
{\epsfxsize3.0in\epsfbox{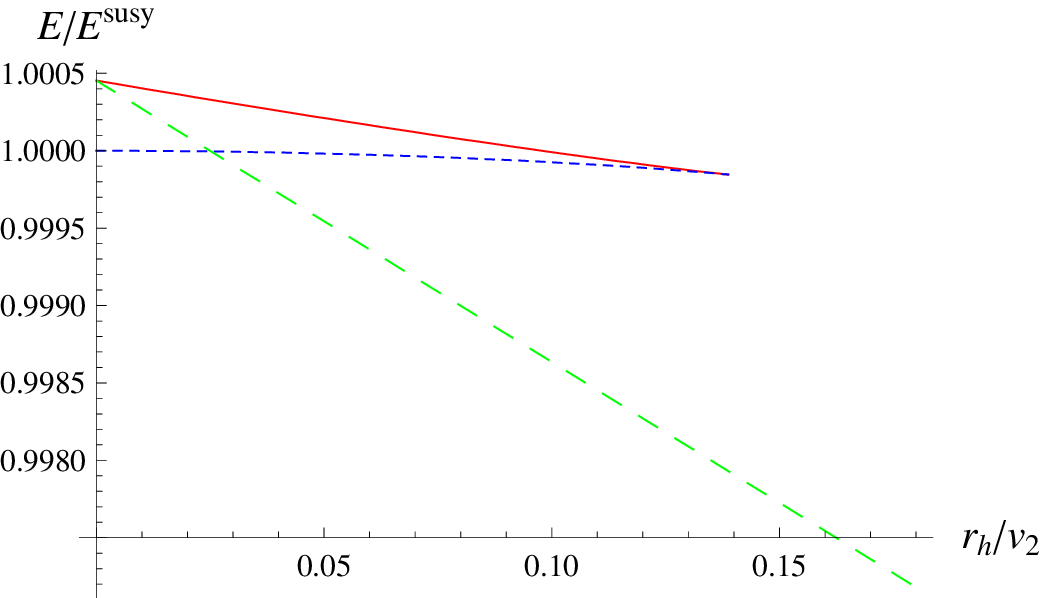}}

So far we discussed the effects of the non-extremality of the $NS5$-branes on the
energetics of $D4$-branes stretched between them. In that discussion we implicitly
assumed that the fivebranes remain static\foot{In field theory language, this is the
assumption that the couplings in the Lagrangian can be taken to be constant in the
early universe. In string theory it is natural for them to be time-dependent.} as the
$D4$-branes move in their background. As mentioned above, this assumption is
generically violated. The gravitational attraction between the non-extremal $NS'$-branes
leads to collapse and merging of the fivebranes on a time scale much shorter than
that associated with the change of $r_h$. In that case, the endpoint of the dynamics
is a system with coincident $NS'$-branes, rather than that of figures 1, 2.

The analysis of energetics above is nevertheless useful, since much of it is applicable
with minor modifications to the generalized ISS system, which will be discussed in the
next subsection and does not suffer from the above instability. It may also be applicable
to the ISS brane system of figures 1, 2 if the Schwarzschild radius $r_h$ decreases
sufficiently fast to avoid the gravitational instability. We do not know why this should
be the case, but for completeness briefly comment next on the early universe dynamics
when the relative motion of the $NS'$-branes can be neglected.

Start at a time where $r_h$ is in the range \RmaxText. In this regime, the only minimum
of the effective action for the $D4$-branes stretched between $NS'$-branes is the
configuration of figure 1. Thus, the $N_f-N_c$ $D4$-branes in figure 2 collapse to the
configuration with unbroken gauge symmetry.

As $r_h$ decreases with time, it enters the regime \RrecText. A local minimum with broken
gauge symmetry appears (short-dashed blue line in figure 5), separated from the symmetric
vacuum by a potential barrier. In that regime, the broken vacuum has higher energy than the
unbroken one (see figure 6). When $r_h$ further decreases below $r_{\rm rec}$, the symmetry
breaking vacuum becomes the global minimum of the potential, but it is still separated from
the symmetric one by a potential barrier. Since in the course of time evolution the system
gets to this regime in the symmetric vacuum, the transition to the true vacuum can only
happen via tunneling. The rate for this process is non-perturbative in $g_s$. Thus, it does
not occur on the time scale associated with the evaporation of the fivebranes, $l_s/g_s^2$.

As $r_h$ decreases further, it gets to the range $r_h < l_s$, where the DBI analysis of
reconnection in appendix B becomes unreliable.\foot{Implicit in the above
discussion is the assumption that $r_{\rm max},r_{\rm rec}\gg l_s$. Otherwise, the DBI
analysis of the energetics becomes unreliable earlier.} At that point, there are two possible
scenarios for the  time evolution. One is that the potential barrier between the two minima
in figure 6 remains finite for all $r_h>0$, and disappears in the limit. In that case, the
tunneling rate increases with time, and eventually the system makes a first order phase
transition to the symmetry breaking vacuum. The second possibility is that the barrier
disappears for a finite value of $r_h$, at which point the system makes a second order phase
transition to the symmetry breaking vacuum. In both cases the endpoint of the dynamics is
as in the field theory analysis \refs{\AbelCR\CraigKX\FischlerXH-\AbelMY}.

Note that throughout the time evolution described above, the supersymmetric vacuum is
separated from the ones studied here by a wide potential barrier. In the above discussion
we implicitly assumed that the system starts at early times in the basin of attraction of the
metastable vacuum. One can ask whether this assumption is natural. We will
discuss this question in the next subsection in the context of the generalized ISS brane
system, where the supersymmetric vacua are visible classically, and the gravitational
instability associated with the separation of the $NS'$-branes is absent.

\subsec{Generalized ISS system}

We start again by recalling what happens in the extremal case, $r_h=0$. The brane
system of figures 3, 4 corresponds to the generalized ISS model \wdef, with the deformation
parameter $\mu_\phi$ determined by the angle $\theta$ via \muphi. It has multiple vacua,
labeled by the integers $k$, $n$ in figure 4. Vacua with $n=0$ are supersymmetric; of the
remaining ones, only those with $k=N_f-N_c$ are locally stable in the field theory regime
(see section 2), and for general values of the geometric parameters in weakly coupled string
theory (section 3).

As indicated in figure 4, for $\theta\not=0$ the $D4$-branes stretched between $NS'$-branes
have a finite extent in $w$. Thus, one needs to be careful in parameterizing their locations
in the $w$-plane. One way of doing that is to label the location of a fourbrane by the
position of its endpoint on the $NS_1'$-brane, $w_1$. The potential for such a $D4$-brane
can again be calculated from the DBI action. Expanding around the origin one finds \GiveonUR
\eqn\Pottwo{ V(w_1) = \frac{T_{4} \Delta y}{g_s} \left( 1 - \frac{a^2 w_1}{w_0} +
\frac{a^2 l_{s}^2  w_{1}^2 }{2 y_{NS}^4}  + \cdots \right)~.}
In \Pottwo, $w_0$ and $w_1$ are taken to be real, although a priori they are arbitrary
complex numbers. The location of the supersymmetric minimum, $w_0$, can be chosen to be
real and positive by a rotation in the $(8,9)$-plane or, in field theory language, by a
phase rotation on the chiral superfields. The potential for the complex scalar field $w_1$
has the property that at the minimum its phase is the same as that of $w_0$. Thus, for
real $w_0$, it is enough to restrict the potential for $w_1$ to a real slice, as we do in \Pottwo.

The potential \Pottwo\ has a minimum at $w_1=w_{\rm min}$,
\eqn\wwmmiinn{w_{\rm min}\simeq{y_{NS}^4\over l_s^2w_0}~.}
As $\theta\to0$, $w_0\to\infty$, and the corresponding potential and minimum approach those
of the ISS system. Since the potential \Pottwo\ is valid only for $w_1\ll |y_{NS}|$, \wwmmiinn\
is valid for
\eqn\thetacon{w_0 \gg \frac{|y_{NS}|^3}{l_{s}^2} ~.}
This implies an upper bound on the deformation parameter $\theta$, \muphi, \intbranes,
\eqn\smalltheta{\theta \ll  \frac{a\Delta y l_s^2}{|y_{NS}|^3}  ~.}
We will restrict to this regime below, for simplicity.

Another constraint on $\theta$ that we will impose is associated with the requirement that all
the vacua in figure 4 be locally stable. As discussed in \refs{\GiveonEW,\GiveonUR}, the lightest
modes of fundamental strings stretched between the color and flavor branes, which correspond in
the field theory to the magnetic quarks denoted by $\varphi$, $\tilde\varphi$ in \metaformphi,
are tachyonic if the distance between the endpoints of these strings on the $NS_1'$-brane, $w_1$,
is smaller than $\sqrt{2\pi a}l_s$. Therefore, in order for all the vacua labeled by $k$, $n$,
in figure 4 to be locally stable, one must have
\eqn\locstab{w_0<{y_{NS}^4\over l_s^3\sqrt{2\pi a}}~.}
The corresponding lower bound on the deformation parameter $\theta$ (or $\mu_\phi$ \muphi)
is easy to satisfy in the regime \thetacon, as is obvious from the fact that the corresponding
bounds on $w_{\rm min}$, (namely $\sqrt{2\pi a} l_s < w_{\rm min} \ll |y_{NS}|$), are compatible.

Our main purpose here is to analyze the effect of non-extremality of the $NS5$-branes on the
energetics of the brane configuration of figure 4. As shown in appendix A, the introduction of
non-zero $\theta$ eliminates the instability of the ISS brane configuration noted above. Since
the two $NS'$-branes are no longer parallel, the attraction between them does not lead to
gravitational collapse. Instead, the fivebranes take a static shape in which each of them develops 
a localized dip towards the other (see figure 7). They also develop a logarithmic shape far from
the intersection. Since we take the extra dimensions  to be non-compact,
this does not present particular difficulties.

To understand the early universe dynamics of the brane system of figure 4, we next study the
energetics of the $D4$-branes in the resulting fivebrane background. We discuss first the
vicinity of the metastable minimum, and then that of the supersymmetric vacuum.

\ifig\loc{Bending of $NS'$-branes.  Axes are offset,  $w \rightarrow w-w_0$, 
$y \rightarrow y-y_1-\half \Delta y$.  This figure was generated using the 
results in appendix A, with  $y_0 = 1$,  $r_h = 0.1$, and $\theta = 0.2$.  
The expansion parameter $r_h/(\theta y_0) = 0.5$, which may be too 
large to be trustworthy, but was chosen to accentuate the features.}
{\epsfxsize3.0in\epsfbox{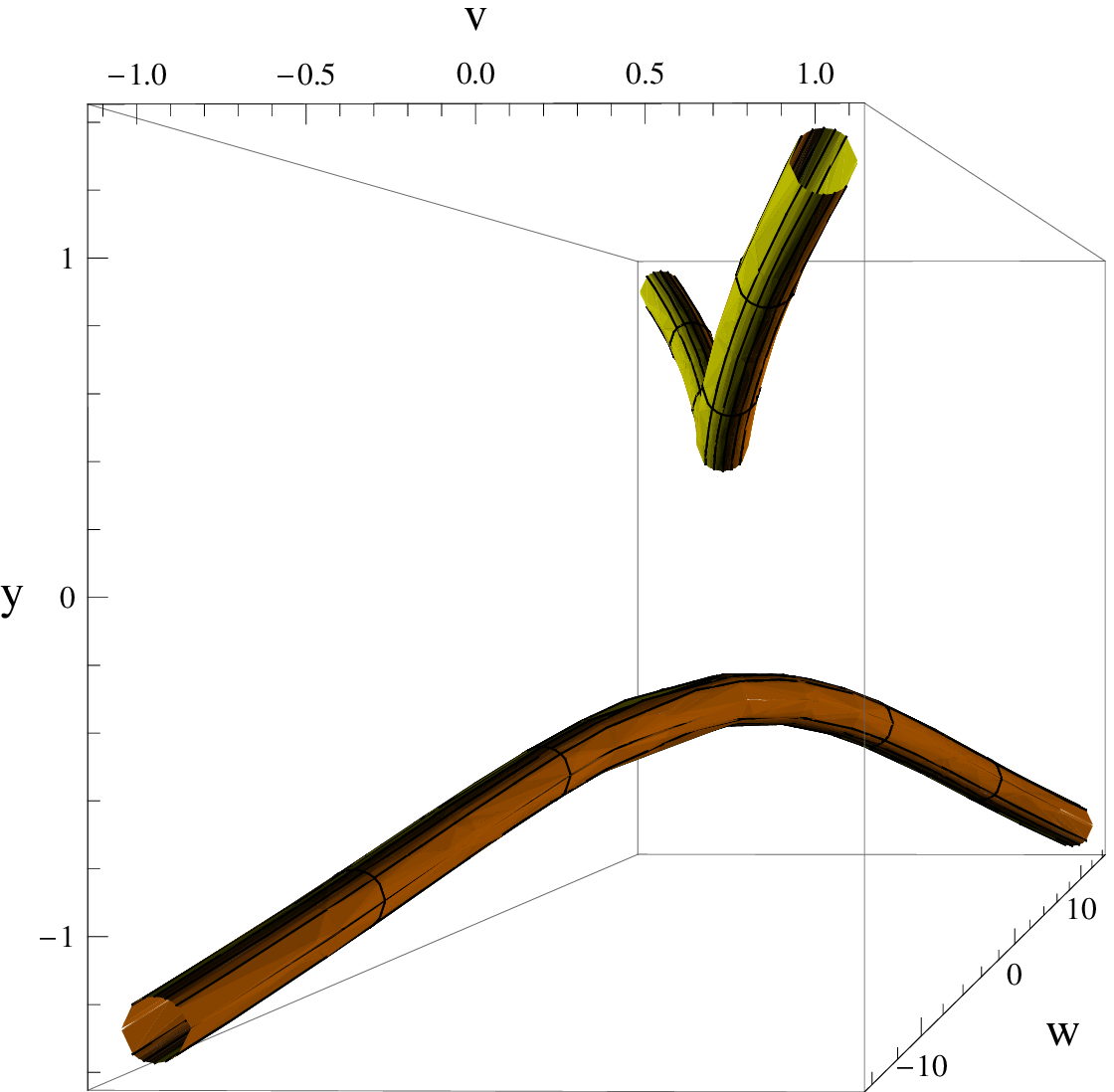}}

For small $w_1$, a generalization of the calculation leading to \Pottwo\ gives
\eqn\Pottworh{ V(w_1) = \frac{T_{4} \Delta y}{g_s} \left[ 1 - \frac{a^2 w_1}{w_0} +
\frac{ (r_{h}^2 +2 a^2 l_{s}^2 ) w_{1}^2 }{4 y_{NS}^4}  + \cdots \right]~.}
The minimum of \Pottworh\ is located at
\eqn\rhmetmin{ w_{\rm min}(r_h) = \frac{w_{\rm min}}{ 1 + r_{h}^2 / 2 a^2 l_{s}^2 } ~,}
with $w_{\rm min}$ given by \wwmmiinn. For small $r_h$ $(r_h\ll al_s)$, $w_{\rm min}(r_h)$ is
close to \wwmmiinn. As $r_h$ increases, $w_{\rm min}$ decreases, and for $r_h\gg al_s$ it
goes like $w_{\rm min}(r_h)\simeq 2a^2y_{NS}^4/w_0r_h^2$.

For $w_1\sim w_0$, \ie\ in the vicinity of the supersymmetric vacuum, one expects on
general grounds the non-extremality of the $NS'$-branes to lead to important effects.
The interaction of the fivebranes depends on two dimensionless parameters, $\theta$ and
$r_h/\Delta y$. As discussed above, for $\theta=0$ and any finite $r_h$ the branes collapse
onto each other and form a single black brane with a larger horizon. On the other hand, for
$r_h=0$ they are mutually BPS and are not deformed at all. In general,  one expects a
smooth interpolation between the two limits.

This problem is discussed in appendix A. As $r_h$ increases, the branes bend
towards each other in a small region near the intersection (see figure 7). The angle 
between them near the intersection also decreases as follows:
\eqn\ParalAng{
\theta_0=\theta\left[1-\left(r_h\over2\theta\Delta y\right)^2+\cdots\right]~.
}
The expansion parameter on which the shape of the fivebranes depends turns out
to be $r_h/\theta\Delta y$. When this parameter becomes of order one, the expansion
breaks down. We expect the horizons of the two black branes to merge around that point,
in a region in $w$ whose size increases with $r_h$. When that happens, any $D4$-branes 
stretching between the $NS'$-branes near $w_0$ disappear into the black hole.

Note that the value of $r_h$ above which $D4$-branes near $w_0$ disappear from the
dynamics, $r_h\sim \theta\Delta y$, is very small. Indeed, using \smalltheta\ one can
show that $\theta\Delta y\ll al_s$. Thus, typically  the supersymmetric minimum at
$w\sim w_0$ is surrounded by a horizon throughout the time evolution of the system.

We are now ready to discuss the early universe dynamics of the branes in the configuration
of figures 3, 4. As before, we will take the Schwarzschild radii of the three $NS5$-branes,
$r_h$, to be the same, and consider the dynamics as $r_h$ decreases. We start at an early
time with $r_h\gg\theta\Delta y$, so that any branes in the basin of attraction of the
supersymmetric vacuum at $w_1\sim w_0$ fall into the  black hole. Hence, all the branes
that survive at late times must be located far from the supersymmetric vacuum, at a relatively
small value of $w$.

To follow the evolution of these branes, we start with $r_h$ in the range \RmaxText, so the
$D4$-branes stretched between the $NS'_2$ and $NS$-branes are unstable to collapse
towards the $NS_1'$-brane. Thus, at this point in the evolution, the system is attracted to a
configuration in which the $N_f-N_c$ color $D4$-branes are stretched between the $NS$
and $NS_1'$-branes, and the $N_f$ flavor branes are stretched between the $NS'$-branes
at a value of $w$ given by \rhmetmin.

As $r_h$ decreases, two things happen:
\item{(1)} The position of the metastable minimum \rhmetmin\ increases.
\item{(2)} The potential barrier that prevents the color and flavor $D4$-branes
from reconnecting decreases (as discussed in the previous subsection).

\noindent
The two effects act in opposite directions -- the first one tends to suppress the reconnection
process, while the second enhances it. For $r_h\gg l_s$ the potential barrier prevents reconnection
from happening. What happens in the stringy regime is not clear, since the DBI analysis of the
reconnection process in appendix B breaks down there. What is clear is that one can choose the
parameters such that the position of the metastable minimum \rhmetmin\ in the stringy regime is
sufficiently large, so that reconnection does not occur as $r_h$ decreases, \ie\
$w_{\rm min}(r_h)>\sqrt{2\pi a} l_s$ for $r_h<l_s$. This is typically the case when the inequality
\locstab\ is satisfied by a wide margin.

It might be that there is a small range of parameters for which $w_{\rm min}$ \wwmmiinn\ is
sufficiently close to $\sqrt{2\pi a}l_s$ that reconnection does occur for some $r_h<l_s$.
Establishing the existence of such a range requires generalizing the analysis of appendix
B beyond the DBI approximation; we will leave this for future work.

Assuming that $w_{\rm min}$ is taken to be large enough that the reconnection of color and flavor
branes does not occur, the brane system approaches at late times the configuration of figure 4
with $k=0$ and $n=N_f$, in which the gauge symmetry is unbroken. This is one of the vacua that
exists in the brane system but not in the field theory of section 2.

If there is a region in parameter space in which the reconnection of color and flavor branes
does occur, the system approaches at late times the configuration with $k=N_f-N_c$, $n=N_c$,
the generalization of the ISS vacuum to non-zero $\mu_\phi$. We stress again, that if such
a region in parameter space exists, it corresponds to a limited range of the parameters, for
which the minimum $w_{\rm min}$ is sufficiently close to the bound $\sqrt{2\pi a}l_s$.

To recapitulate, we find that if the parameters of the generalized ISS brane system are chosen to
be in the range in which all vacua in figure 4 are locally stable, early universe dynamics typically
(\ie\ for generic values of the parameters and initial conditions) leads the system to the most
stringy and most symmetric of these vacua. One can fine-tune the parameters and/or initial conditions
so that the system approaches one of the other vacua, but this is not generic.

\newsec{Discussion}

The work reported in this paper is part of a long term program of building phenomenologically 
viable models of nature in string theory, using configurations of intersecting $D$-branes and
$NS5$-branes. The first steps in this program were taken in the late 1990's when it was shown 
that such configurations, which arise naturally in string theory, provide a simple way to realize 
supersymmetric gauge theories of the sort expected to play a role in nature. An added bonus 
of the string theory realizations of these theories was an improved understanding of non-trivial 
dynamical phenomena such as mirror symmetry, Seiberg duality and Seiberg-Witten curves 
(see \GiveonSR\ for a review). 

In the last few years it became clear that these brane systems are also very useful for studying 
supersymmetry breaking. They exhibit metastable vacua of the sort discovered in the corresponding
gauge theories in \IntriligatorDD. Moreover, in some cases, such as the generalized ISS model, 
they have a much richer vacuum structure. The additional metastable vacua present in the brane
systems but not in the corresponding gauge theories are in fact phenomenologically appealing
for describing supersymmetry breaking in nature.

In this paper we addressed the early universe dynamics of these brane systems. We found that
they provide a picture which is closely related to the corresponding studies in field theory. The analog 
of turning on a finite temperature in the field theory analysis is replacing the $NS5$-branes by
non-extremal black branes with a finite Schwarzschild radius $r_h$. The analog of the finite temperature
second order phase transition associated with condensation of the magnetic quarks $q$, $\tilde q$
is the transition associated with reconnection of the $D4$-branes (discussed around figure 5).
The analog of the destabilization of the supersymmetric vacuum by thermal effects discussed in
\AbelCR\ is the merging of the non-extremal $NS'$-branes at an angle $\theta$ into a localized
black hole that swallows any $D4$-branes stretched between the $NS'$-branes near the intersection
of the two.

The string theory realization provides again some unexpected bonuses:
\item{(1)} R-symmetry breaking, which in the field theory discussions is 
important for phenomenological reasons, must be introduced in string 
theory for dynamical reasons. Indeed, the original ISS system 
\refs{\OoguriBG\FrancoHT\BenaRG-\GiveonFK}, which has unbroken
R-symmetry, is an unnatural endpoint of early universe dynamics, due to the
attraction of the $NS'$-branes to each other.\foot{The brane construction in which 
the $NS_2'$-brane is replaced by $N_f$ $D6$-branes suffers from the same problem.}
\item{(2)} As mentioned above, the generalized ISS system, which has broken R-symmetry,
has many metastable vacua that do not exist in field theory and are phenomenologically 
appealing. Interestingly, early universe dynamics drives the system to the most stringy 
of these vacua. This vacuum has the largest unbroken gauge symmetry, but at the same
time it is the least stable of all the vacua labeled by $(n,k)$ in figure 4. It is an interesting
candidate for the supersymmetry breaking sector of a model of nature.

\noindent
A natural next step in the program described above is to use the brane models to realize 
inflation in string theory, and connect the inflationary period to the dynamics studied here 
(which we imagine to be post-inflationary). It would also be interesting to couple these 
models to a sector that realizes the MSSM in string theory and study the experimental 
consequences of the resulting models. For both purposes it is important to make the extra 
dimensions compact.

\bigskip

\noindent{\bf Acknowledgements:} We thank A. Giveon, M. Porrati, S. J. Rey and R. Wald  for 
discussions. DK thanks the Weizmann Institute and the organizers of the Fifth Crete Regional 
Meeting in String Theory for hospitality during part of this work. JM would like to thank the 
organizers  of the New Perspectives in String Theory workshop at the Galileo Galilei Institute for 
Theoretical Physics and Simons Workshop in Mathematics and Physics 2009 for hospitality 
during part of this work. DK and OL are supported in part by DOE grant DE-FG02-90ER40560 
and the National Science Foundation under Grant 0529954. JM is supported in part by a 
Ledley fellowship. AR is supported in part by a Department of Education GAANN fellowship
P200A060226.

\appendix{A}{Interaction between non-extremal fivebranes}

On several occasions in section 4, we encountered an $NS5$-brane in the geometry produced by another non-extremal fivebrane.  In this appendix we will analyze the profile of an $NS5$-brane caused by such a geometry; in particular, we will derive equations \NSsmallw\ and \ParalAng.  Strictly speaking, this analysis is valid for a probe extremal fivebrane in the background of a non-extremal one.  However, one expects that the results are applicable to the case when both fivebranes are non-extremal, provided that their separation is much greater than their horizon radii.

\subsec{Linear analysis for a general angle}

We begin with the background of a non-extremal $NS5$-brane, that wraps the $\IR^{3,1}$ labeled by $(x^i,t)$ and the $v$-plane, and sits at $y = x^7 = w = 0$.  This is equations \nonex\ and \fH\ in the text\foot{Except that here we put the source at $y=0$ to simplify notation.}, which we rewrite here in a more convenient fashion:
\eqn\nonexnew{ \eqalign{ & ds^2 = - f dt^2 + H \left[ \left( f^{-1} -1 \right) dr^2 + dy^2 + (dx^7)^2 + dw d\wbar \right] +
(dx^i)^2 + dv d \vbar~, \cr
& e^{2(\varphi-\varphi_0)} = H(r) ~,}}
with $r^2 = y^2 + (x^7)^2 + |w|^2$ and
\eqn\fHnew{ f(r) = 1 - \frac{r_h^2}{r^2} ~, ~~~~~ H(r) = 1 + \frac{l_{s}^2}{r^2}~.}
Let us place a probe fivebrane in the background \nonexnew.  The probe brane wraps the $\IR^{3,1}$ labeled by $(x^i,t)$, and when $r_h = 0$, it follows a straight (complex) line:
\eqn\ProbeExtr{ v=\alpha w~,\quad y=y_0~,\quad x_7=0~.}
For small $r_h$ the leading correction to this profile can be written as\foot{Here we are looking for static configurations.}
\eqn\ProbeNExtr{ v = \alpha w+r_h^2{\tilde v}(w,\wbar)~, \quad  y = y_0+r_h^2 {\tilde y} (w,\wbar)~, \quad x_7 = 0~.}
In the following analysis we will restrict to the regime $r_h \gg l_s$ so that we may set $H=1$ in \nonexnew.

The dynamics of the probe fivebrane is governed by the Nambu--Goto type action,
\eqn\ApNSact{ \SS = -T_{NS} \int d^6 \sigma e^{-2\varphi} \sqrt{-{\rm det} P(G)_{ab}}~,}
where $P(G)_{ab}$ is the metric induced on the worldvolume of the brane.  In general \ApNSact\ should contain terms involving the Neveu-Schwarz $B$-field: $P(G)$ should be generalized to $P(G+B)$ in the Nambu-Goto term, and there should be a Wess-Zumino term coupling the potential $B_6$, where $d B_6 = \star dB$, to the worldvolume of the probe brane.  The contribution of the $B$-field to the Nambu-Goto term vanishes on the embedding \ProbeNExtr, while the Wess-Zumino term is in general nonzero, but vanishes in the regime $r_h\gg l_s$ we work in.  Expanding \ApNSact\ in the powers of $r_h$, we arrive at linear 
equations for ${\tilde y}$ and ${\tilde v}$:
\eqn\TimeEOM{ \eqalign{ y_0 \left[(1+\alpha^2) w \wbar + (\alpha^2-1) y_0^2 \right] - 4 (y_0^2 + w \wbar)^3 \d_w \d_{\wbar} {\tilde y} = 0~, \cr
\alpha w \left[ (1+\alpha^2) w \wbar + (\alpha^2-1) y_0^2 \right]+ 4 ( y_0^2 + w \wbar)^3 \d_w \d_{\wbar} {\tilde v} = 0~.}}
These equations have a unique solution which is regular at $w=0$ and has ${\tilde y}(0)={\tilde v}(0)=0$, and it leads to the following expressions for $y$ and $v$:
\eqn\NoTimeSolBB{ \eqalign{ & y = y_0 - \frac{r_h^2 |w|^2 }{ 4 y_0 (y_0^2 + |w|^2)} + \frac{r_h^2 \alpha^2}{4 y_0} \ln{ \left( \frac{y_0^2 + |w|^2}{y_0^2} \right)} ~,\cr
& v = \alpha w \left\{ 1+ \frac{r_h^2}{4 |w|^2} \left[ -\frac{|w|^2}{y_0^2 + |w|^2} + (1+\alpha^2) \ln{ \left( \frac{ y_0^2 + |w|^2}{y_0^2} \right)} \right] \right\} ~.}}

The properties of this solution at small and large values of $\alpha$ will be discussed in the next two subsections; here we make a few comments regarding qualitative features of the general solution.
\item{(1)}  Consider the small $w$ behavior of the solution for $y$.  Expanding around $w = 0$, we find
\eqn\bumpordip{ y = y_0 \left[ 1 + (\alpha^2 -1) \frac{ r_{h}^2 |w|^2}{4 y_{0}^4} + \OO\left( \frac{|w|^4}{y_{0}^4} \right) \right]~.}
When $\alpha <1$, corresponding to an asymptotic angle between the branes greater than $45^\circ$, the probe brane bulges out near $w = 0$ and bends around the non-extremal brane.  When $\alpha >1$, so that the asymptotic angle is less than $45^{\circ}$, the probe brane dips down towards the non-extremal brane.  (When $\alpha =1$, the next term in the expansion \bumpordip\ shows that the probe brane dips down in this case as well).
\item{(2)}  While $v$ approaches the straight line $v = \alpha w$ as $|w| \rightarrow \infty$, $y$ on the other hand has a logarithmic running in $|w|$ (unless $\alpha = 0$).  However, this behavior is only noticeably different from an asymptotically constant profile on distance scales much larger than those we consider in this paper.
\item{(3)}  When both branes are non-extremal with equal horizon radii, each brane will of course have an equal and opposite effect on the other.  In this case, \NoTimeSolBB\ accounts for the net effect, to leading order in $r_{h}^2/y_{0}^2$.

\subsec{$NS'$-branes in the background of $NS$: $\alpha=0$}

To describe the interaction between the $NS$-brane and $NS'$-branes, we set $\alpha = 0$ in the above analysis.  Then the role of the probe brane is played by the $NS'$-brane, which sits at $v = 0$ (or a constant), and stretches in $w$.  Its profile in $y$ is given by
\eqn\SolBBZer{ y = y_0 - \frac{r_h^2 |w|^2 }{4 y_0( y_0^2+ |w|^2)}~.}
Moreover, since the configuration with $\alpha=0$ has a symmetry $v \leftrightarrow -v$, the branes should remain orthogonal to all orders in $r_h$.

The value $\alpha=0$ is rather special: unlike the general result \NoTimeSolBB, equation \SolBBZer\ does not have a logarithmic running at large values of $w$. Then, assuming the distance between branes in the $y$-direction, $L$, is measured at infinity, we find the relation\foot{Here, and in the following, we drop higher order terms in $r_{h}^2/y_{0}^2$ and $r_{h}^2/L^2$.}
\eqn\ApSolAA{ L = y_0 - \frac{r_h^2}{4y_0}~,}
which leads to a modified version of \SolBBZer,
\eqn\SolBBOne{ y = L \left[1+\frac{r_h^2}{4( L^2 + |w|^2)} \right]~.}
To bring this expression into the form \NSsmallw, we recall that $L=y_i-y_{NS}$ (where $i=1,2$ for $NS_{1}'$, $NS_{2}'$ respectively), and shift the origin of the $y$-direction so that the $NS$-brane sits at $y = y_{NS}$:
\eqn\textresone{ y = y_i + (y_i - y_{NS}) \frac{ r_{h}^2}{4( ( y_i - y_{NS})^2 + |w|^2 )}~.}

\subsec{Intersection of $NS_{1}'$ and $NS_{2}'$: large $\alpha$}

To study the interaction of the $NS'$-branes with each other, we first interchange $v$ and $w$ in \nonexnew, so that the background describes the geometry produced by the $NS_{1}'$-brane:
\eqn\nonexNSone{ \eqalign{ & ds^2 = - f dt^2 + H \left[ \left( f^{-1} -1 \right) d{\tilde r}^2 + dy^2 + (dx^7)^2 + dv d\vbar \right] +
(dx^i)^2 + dw d \wbar~, \cr
& e^{2(\varphi-\varphi_0)} = H({\tilde r}) ~,}}
with ${\tilde r}^2 = y^2 + (x^7)^2 + |v|^2$.  If we interchange the role of $v$ and $w$ in \ProbeExtr\ and \ProbeNExtr\ as well, we can describe the $NS_{2}'$ profile by identifying $\alpha \equiv \cot{\theta}$:
\eqn\ProbeNStwo{ w = v \cot{\theta} + r_h^2{\tilde w}(v,\vbar)~, \quad  y = y_0+r_h^2 {\tilde y} (v,\vbar)~, \quad x_7 = 0~.}
Then it is clear that the general solution to the linearized equations of motion can be obtained from \NoTimeSolBB\ by making the replacements $v \leftrightarrow w$, $\alpha \rightarrow \cot{\theta}$:
\eqn\NoTimeSolCC{ \eqalign{ & y = y_0 - \frac{r_h^2 |v|^2 }{ 4 y_0 (y_0^2 + |v|^2)} + \frac{r_h^2 \cot^2{\theta} }{4 y_0} \ln{ \left( \frac{y_0^2 + |v|^2}{y_0^2} \right)} ~,\cr
& w =  v \cot{\theta} \left\{ 1+ \frac{r_h^2}{4 |v|^2} \left[ -\frac{|v|^2}{y_0^2 + |v|^2} + \csc^2{\theta} \ln{ \left( \frac{ y_0^2 + |v|^2}{y_0^2} \right)} \right] \right\} ~.}}

We are interested in very small values of $\theta$, according to \smalltheta.  \NoTimeSolCC\ shows that $y$ has a logarithmic behavior at large $v$, but we can define a separation between branes $\Delta y$ by truncating this running at some large value of $v$,
\eqn\NotSerious{ \Delta y \equiv y_0 - \frac{r_h^2}{4y_0} + \frac{r_h^2}{\theta^2 y_0} c_1\approx y_0 + \frac{r_h^2}{\theta^2 y_0} c_1~,}
where $c_1$ is some constant of order one. As the non-extremality parameter $r_h$ increases, while the separation between branes $\Delta y$ is kept fixed, the minimal distance between branes $y_0$ decreases.  Inverting \NotSerious, we find
\eqn\NotSeriousInv{ y_0 \approx \Delta y - \frac{r_h^2}{\theta^2 \Delta y} c_1~.}
As the value of $r_h$ increases, the branes are also becoming more parallel.  The expression \NoTimeSolCC\ for $w$ shows that, while the angle between the branes approaches $\theta$ at large values of $v$, at $v=0$ one finds a smaller angle $\theta_0$:
\eqn\ApThZer{ \frac{1}{\theta_0} \approx \frac{1}{\theta} \left[1+\frac{r_h^2}{4y_0^2 \theta^2} \right]~.}
Using \NotSeriousInv\ in \ApThZer, we can reproduce \ParalAng.

If the minimal distance between branes, $y_0$, becomes of order $r_h$, then the horizons of the $NS_{1}'$ and $NS_{2}'$-branes touch, and these objects merge into a single black brane.  Unfortunately the relation \NotSerious\ breaks down long before this point.  \NotSerious\ follows from \NoTimeSolCC, which is based on our linear analysis.  In order for \NoTimeSolCC, \NotSerious\ to be valid, we must require that $r_h/(\theta y_0) \ll 1$.  Looking at the higher-order contributions to \NotSerious, one finds a general expansion
\eqn\AlmostSerious{ \Delta y = y_0 \left[ 1+ \sum_{n=1}^\infty c_n  
\left(\frac{r_h}{\theta y_0}\right)^{2n} \right]~.}
Assuming that this expansion is invertible,
\eqn\AlmostSeriousInv{ y_0 = \Delta y \left[ 1+ \sum_{n=1}^\infty 
d_n \left( \frac{r_h}{\theta\Delta y}\right)^{2n} \right] = \Delta y~ g \left(\frac{r_h}{\theta\Delta y}\right)~,}
we conclude that the branes connect if the function $g$ takes a value $r_h/(\Delta y)\ll 1$.  For a smooth function $g$ this happens close to the point where $g(x)$ vanishes, and this leads to the expression for the critical non-extremality parameter $(r_h)_{merge}$:
\eqn\rmerge{ (r_h)_{merge} = x_0 \theta\Delta y~, \qquad g(x_0) = 0~.}
For $r_h>(r_h)_{merge}$, the $NS_{1}'$- and $NS_{2}'$-branes merge in some region close to $v=0$.

To conclude this subsection, we observe that the critical value of $r_h$ given by \AlmostSeriousInv\ is equal to zero for parallel branes. This is consistent with the fact that such branes experience a finite force per unit length, so for any nonzero value of the non-extremality parameter they merge in finite time, and no static solution describing separated branes exists.

\appendix{B}{$D4$-brane dynamics in the non--extremal $NS_{1}'$ background}

In section 4.2 we discussed the dynamics of $D4$-branes in the geometry produced by a non-extremal $NS_{1}'$, and found that the $D$-branes had an interesting vacuum structure if the non-extremality parameter, $r_h$, is sufficiently large.  In this appendix we will present the details of the analysis which led to this conclusion.

The $NS_{1}'$-brane geometry is given by
\eqn\nonexPrm{\eqalign{ &ds^2 = -f({\tilde r}) dt^2+H({\tilde r}) (f({\tilde r})^{-1} d{\tilde r}^2 + {\tilde r}^2 d{\tilde\Omega}_3^2) + (dx^i)^2 + dwd{\bar w}~, \cr
& e^{2(\varphi-\varphi_0)} = H({\tilde r})~.}}
Here $({\tilde r},{\tilde\Omega}_3)$ are spherical coordinates on the four-dimensional space transverse to the brane\foot{We introduce tildes in this expression to distinguish the spherical coordinates centered on the $NS_{1}'$ from those centered on the $NS$. The latter were used in the beginning of section 4.}; in particular, ${\tilde r}=\sqrt{(y-y_1)^2+(x^7)^2+v{\bar v}}$.  The functions $f$, $H$ are given by
\eqn\eqnfHPrm{ f({\tilde r}) = 1-\frac{r_h^2}{{\tilde r}^2}~,\quad H({\tilde r}) = 1+\frac{l_s^2}{{\tilde r}^2}~.}

Starting with the geometry \nonexPrm, we consider a $D4$-brane which is stretched between the $NS$- and $NS_{2}'$-branes and extended in the directions $(x^i)$.  Assuming that the $D4$-brane is attached to the $NS5$-branes at points $(v,w,y)=(0,0,y_{NS})$ and $(v,w,y)=(v_2,0,y_2)$, we conclude that the $D4$-brane is localized at $(w,{\rm Im}(v))=(0,0)$, and its profile is described by $y=y({\rm Re}(v))$.  It is convenient to introduce polar coordinates $(\rho,\phi)$ in the plane spanned by $({\rm Re}(v),y)$,
\eqn\vyrhophi{ {\rm Re}(v)+i(y-y_1) = \rho e^{i\phi}~,}
and to parameterize the worldvolume of the $D4$-brane in terms of $(x^i,t)$ and the radial coordinate $\rho$.  The profile of the brane in the
${\rm Re}(v)$-$y$ plane, $\phi(\rho)$, can be found by minimizing the DBI action,
\eqn\ActAA{ \eqalign{  S & = -T_4\int d^4 x d\rho e^{-\varphi}\sqrt{-{\rm det}P(G+B)_{ab}} \cr
& = -\frac{T_4}{g}\int d^4x \int d\rho \sqrt{1+\rho^2 f(\rho)(\d_\rho\phi)^2}~.}}
The limits of integration in the $\rho$-coordinate will de determined below.  The action \ActAA\ leads to the equation of motion for $\phi(\rho)$,
\eqn\aaaO{ \d_\rho \left[ \frac{\rho^2 f(\rho)\d_\rho\phi}{\sqrt{1+\rho^2 f(\rho)(\d_\rho\phi)^2}} \right] = 0~,}
and integrating this equation we determine the location of the $D4$-brane in a plane parameterized by
the polar coordinates $(\rho,\phi)$:
\eqn\aaaT{ \phi_\pm (\rho) = \pm\int_{\rho_0}^\rho \frac{\sqrt{\rho_0^2-r_h^2} dr}{\sqrt{(\rho^2-r_h^2)(\rho^2-\rho_0^2)}}+\phi_0~.}
The integration constants, $(\rho_0,\phi_0)$, specify the location of the point where the distance between $D4$- and $NS_{1}'$-branes reaches the minimal value, $\rho_0$.  Although the integral entering \aaaT\  can be evaluated in terms of elliptic integrals $F$ and $K$,
\eqn\exactint{ \phi_\pm(\rho) = \pm\frac{\sqrt{ \rho_{0}^2 - r_{h}^2 }}{\rho_0} \left\{ K\left(\frac{r_{h}^2}{\rho_{0}^2 }\right) - F\left(\arcsin\frac{\rho_0}{\rho},\frac{r_{h}^2}{\rho_{0}^2}\right) \right\} + \phi_0~,}
it is convenient to work directly with \aaaT.

The integration constants, $(\rho_0, \phi_0)$, must be determined in terms of the boundary data, \ie\ the location of the endpoints of the $D4$-brane.  In the setup depicted in figure 5, the $D4$-brane is stretched between two $NS5$-branes, and its endpoints are located at\foot{For now we ignore a possible sliding of the endpoint of the $D4$-brane that is attached to the $NS$-brane.  We will come back to this effect at the end of the appendix.}
\eqn\EndPoint{ (\rho_1,\phi_1)=(y_1-y_{NS},-\frac{\pi}{2})~,\qquad (\rho_2,\phi_2)=(\sqrt{(\Delta y)^2+v_2^2}, \frac{\pi}{2}-\arctan\frac{v_2}{\Delta y})~.}
One solution, satisfying the boundary conditions, consists of two disconnected components, each stretching radially outward from the horizon to the endpoints $(\rho_1, \phi_1)$ or $(\rho_2,\phi_2)$ respectively.  This solution has $\partial_\rho \phi = 0$ along the entire trajectory, corresponding to $\rho_0 = r_h$ in \aaaT. We refer to this configuration as the ``disconnected'' solution, and it is represented by the green long-dashed line in figure 5.

To look for connected solutions with $\rho_0 > r_h$, we proceed as follows.  Assuming that $NS_{1}'$ is near-extremal and the parameter $a$ introduced in \adef\ is small,
\eqn\aaaR{ \frac{r_h}{\rho_1} \ll 1,\quad \frac{r_h}{\rho_2} \ll 1,\quad \pi-\phi_2+\phi_1\ll 1~,}
we can simplify the integral appearing in \aaaT.  It is easy to see that the inequalities \aaaR\  imply that $\phi_0$ is close to zero, and we begin by evaluating $\phi_2-\phi_0$, which is close to $\frac{\pi}{2}$.  Taking the plus sign in \aaaT\ and shifting $\phi$, we define
\eqn\PhiHat{ {\hat\phi}_+(\rho) \equiv \phi_+ (\rho)-\phi_0 = \int_{\rho_0}^\rho \frac{\sqrt{\rho_0^2-r_h^2} d\rho}{\sqrt{(\rho^2-r_h^2)(\rho^2-\rho_0^2)}}~.}
To analyze this expression, it is convenient to introduce dimensionless quantities $\xi$ and $\kappa$:
\eqn\aaaFone{ \xi(\rho) = \sqrt{\frac{\rho^2-\rho_0^2}{\rho_0^2-r_h^2}}~, \qquad \kappa = \frac{r_h^2}{\rho_0^2-r_h^2}~.}
Rewriting equation \PhiHat\ in terms of these variables, we find
\eqn\aaaF{ {\hat\phi}_+(\rho) = \int_0^{\xi(\rho)} \frac{d\xi}{\sqrt{(\xi^2+1)(\xi^2+1+\kappa)}}~.}
Notice that ${\hat\phi}_+(\rho)$ is an increasing function of $\rho$ and a decreasing function of $\kappa$. This function reaches its maximal value at $\rho=\infty$ and $\kappa=0$ (which corresponds to $\rho_0=\infty$),
\eqn\phimaxinf{ {\hat\phi}_+^{max}(\infty) = \int_0^{\infty} \frac{d\xi}{\xi^2+1} = \frac{\pi}{2}~.}
If the $D4$-brane begins at a finite $\rho=\rho_2$, then there are upper bounds on $\rho$ and $\rho_0$, which decrease the maximal angle
${\hat\phi}_+^{max}$. To give  ${\hat\phi}_+^{max}$ which is close to $\pi/2$, both $\kappa$ and $\rho_0/\rho_2$ must be small; this requirement leads to the following inequalities:
\eqn\aaAprx{ r_h \ll \rho_0 \ll \rho_2~.}
Expanding \aaaF\  for small $\kappa$, we find
\eqn\PhiOne{ {\hat\phi}_+(\rho) = \arctan \xi(\rho) - \frac{\kappa}{4} \left[\frac{\xi(\rho)}{1+\xi^2(\rho)} + \arctan \xi(\rho) \right] + O(\kappa^2)~.}
In the approximation \aaAprx, \PhiOne\ simplifies to
\eqn\PhiPlus{ \phi_+(\rho_2) = \phi_0+\frac{\pi}{2}-\frac{\rho_0}{\rho_2}-\frac{\pi}{8} \frac{r_h^2}{\rho_0^2} + O\left( \frac{\rho^3_0}{\rho^3_2} \right) + O\left( \frac{r_h^4}{\rho_0^4} \right)~.}
Similar manipulations lead to
\eqn\PhiMinus{ \phi_-(\rho_1) = \phi_0 - \left[ \frac{\pi}{2} - \frac{\rho_0}{\rho_1} - \frac{\pi}{8}
\frac{r_h^2}{\rho_0^2}+ O\left( \frac{\rho^3_0}{\rho^3_1} \right) +O\left( \frac{r_h^4}{\rho_0^4} \right) \right]~,}
and combination of these two expressions gives the angle between the endpoints of the $D4$-brane:
\eqn\TotalPhi{ \arctan a = \pi - (\phi_+ - \phi_-) \simeq \frac{\rho_0}{\rho_1} + \frac{\rho_0}{\rho_2} + \frac{\pi}{4}\frac{r_h^2}{\rho_0^2}~.}

Equation \TotalPhi\ can be used to determine $\rho_0$ for the given values of $(r_h,\rho_1,\rho_2,a)$. The solution does not always exist, and when it does, it is not unique. First we notice that the right hand side of \TotalPhi\  is bounded from below:
\eqn\RhoCrit{ \frac{\rho_0}{\rho_1} + \frac{\rho_0}{\rho_2} + \frac{\pi}{4}\frac{r_h^2}{\rho_0^2} \ge \frac{3\pi}{4} \frac{r_h^2}{\rho_c^2}~, \qquad \rho_c^3 \equiv\frac{\pi}{2} \frac{\rho_1\rho_2r_h^2}{\rho_1+\rho_2}~.}
The critical value $\rho_0=\rho_c$ corresponds to the minimal angle\foot{Here and below we replace $\arctan a$ by $a$, which is assumed to be small.},
\eqn \ThetaMin{ a_{min} = \frac{3\pi}{4}\frac{r_h^2}{\rho_c^2}~,}
for which \TotalPhi\  has a solution. Alternatively, one can fix $a$, $\rho_1$ and $\rho_2$; then \ThetaMin\  can be interpreted as a relation for the maximal deviation from extremality which still allows for a connected solution:
\eqn \RhMax{ (r_h)_{max} = \left(\frac{16a^3}{27\pi}\right)^{1/2} \frac{\rho_1\rho_2}{\rho_1+\rho_2}~.}
If $a < a_{min}$ (or $r_h > (r_h)_{max}$) then there is no (connected) brane configuration satisfying the boundary conditions.

For $a>a_{min}$ (or $r_h < (r_h)_{max}$) equation \TotalPhi\  has multiple solutions, and to analyze them, it is convenient to rewrite \TotalPhi\  in terms of
$\rho_c$ and $a_{min}$:
\eqn\TotalPhiRed{ \frac{1}{3} - \frac{a}{a_{min}} \left( \frac{\rho_0}{\rho_c} \right)^2+\frac{2}{3} \left( \frac{\rho_0}{\rho_c} \right)^3 = 0~.}
One of the solutions of this cubic equation has negative $\rho_0$ and is not physical.  Now we compare the energy of the two physical solutions with the energy of the disconnected straight branes.

Starting with a general expression for the energy density which follows from the action \ActAA,
\eqn\EnrgDens{ {\cal E} = \frac{T_4}{g_s} \int d\rho \sqrt{1+\rho^2 f(\d_\rho{\phi})^2}~,}
and substituting the solution \aaaT\ for $\phi(\rho)$, we find the energy of the curved $D4$-brane as a function of the minimal distance $\rho_0$:
\eqn\EnrgOne{ {\cal E} = \frac{T_4}{g_s} \int_{\rho_0}^{\rho_1} d\rho \left( \frac{\rho^2-r_h^2}{\rho^2-\rho_0^2} \right)^{1/2} + \frac{T_4}{g_s} \int_{\rho_0}^{\rho_2} d\rho \left( \frac{\rho^2-r_h^2}{\rho^2-\rho_0^2} \right)^{1/2}~.}
The straight-brane disconnected solution has $\rho_0 = r_h$ and energy
\eqn\EnrgZero{ {\cal E}_{straight} = \frac{T_4}{g_s} \int_{r_h}^{\rho_1} d\rho + \frac{T_4}{g_s} \int_{r_h}^{\rho_2} d\rho = \frac{T_4}{g_s} (\rho_1-r_h+\rho_2-r_h)~.}
Subtracting this expression from \EnrgOne , we find
\eqn\DltOne{\eqalign{ \Delta{\cal E} = ~& {\cal E}-{\cal E}_{straight} = \frac{T_4}{g_s} \left\{r_h - \rho_0 + \int_{\rho_0}^{\rho_1} d\rho \left[ \left( \frac{\rho^2-r_h^2}{\rho^2-\rho_0^2} \right)^{1/2} - 1 \right] \right\} + \{\rho_1\rightarrow \rho_2\} \cr
= ~ & \frac{T_4}{g_s} \left\{ r_h - \rho_0 + \frac{ \rho_0^2-r_h^2}{\rho_0} \int_{1}^{\rho_1/\rho_0} \frac{du}{\sqrt{u^2-1} (\sqrt{u^2-1} + \sqrt{u^2 - (r_h/\rho_0)^2 })}\right\} +  \cr
& + \{\rho_1\rightarrow \rho_2\}~.}}
Eventually we want to look at $\Delta {\cal E}$ as a function of $(\rho_1,\rho_2,r_h,a)$, but we postpone the substitution of $\rho_0=\rho_0(\rho_1,\rho_2,r_h,a)$ into \DltOne\ until a later stage.

The connected minimum is favorable when $\Delta{\cal E}<0$; this implies
\eqn\bbbO{ \left(1+\frac{r_h}{\rho_0}\right) \int_{1}^{(\frac{\rho_1}{r_h})\cdot (\frac{r_h}{\rho_0})} \frac{du}{\sqrt{u^2-1} (\sqrt{u^2-1} + \sqrt{u^2-(r_h/\rho_0)^2 })} + \{ \rho_1 \rightarrow \rho_2 \} < 2~.}
The left-hand side of this expression is a monotonically-increasing function of $r_h/\rho_0$ (while $\rho_1/r_h$ and $\rho_2/r_h$ are kept fixed), so it can be equal to $2$ for only one value of $\rho_0$. The explicit  construction will demonstrate that this value satisfies the inequalities
\eqn \bbbOp{ r_h \ll \rho_0 \ll \rho_1, \rho_2~.}
Assuming that $\rho_0\gg r_h$ and keeping only linear terms in $r_h/\rho_0$, we can evaluate the integrals in the left--hand side of \bbbO:
\eqn \bbbT{ \left(1+\frac{r_h}{\rho_0} \right) \left\{ \left[\sqrt{u^2-1}-u\right]_1^{\rho_1/\rho_0} + \left[ \sqrt{u^2-1}-u \right]_1^{\rho_2/\rho_0} \right\}< 2~.}
In the regime \bbbOp, we can expand \bbbT\ in small parameters and find the condition
\eqn \bbbR{ \rho_0^2 > \frac{4\rho_1 \rho_2 r_h}{\rho_1+\rho_2} = \frac{8}{\pi}\frac{\rho_c^3}{r_h}~,}
which justifies approximation $\rho_0\gg r_h$.  For future reference we also write an approximate expression for $\Delta{\cal E}$ which is valid in the regime \bbbOp:
\eqn\DltOneAppr{ \Delta{\cal E} = \frac{T_4}{g_s} (\rho_0-r_h) \left\{ \frac{2r_h}{\rho_0} - \frac{\rho_0}{2\rho_1} - \frac{\rho_0}{2\rho_2} \right\} = \frac{T_4}{g_s} \frac{2r_h}{\rho_0} (\rho_0-r_h) \left\{ 1 - \frac{\pi}{8} \frac{\rho^2_0r_h}{\rho_c^3} \right\}~.}

We found that a connected configuration characterized by $\rho_0$ is energetically preferred over the disconnected configuration as long as  \bbbR\  is satisfied. It is convenient to rewrite this relation in terms of the geometric data (such as $a,r_h,\rho_1,\rho_2$) rather than $\rho_0$, which is determined dynamically.  Substituting \bbbR\  into \TotalPhi, we find that reconnection is energetically preferred if $a$ is larger than a critical value $a_{rec}$,
\eqn \ThetaBrec{ a > a_{rec} \equiv 2\sqrt{\frac{r_h(\rho_1+\rho_2)}{\rho_1\rho_2}} = \frac{4\sqrt{2}}{3\sqrt{\pi}} \sqrt{\frac{\rho_c}{r_h}} ~ a_{min}~.}
For angles satisfying this inequality (so that $a \gg a_{min}$), equation \TotalPhiRed\  has two solutions with positive $\rho_0$,
\eqn\ImpactA{ \rho_0 \simeq \rho_c \frac{3a}{2a_{min}} = \rho_c \sqrt{\frac{8\rho_c}{\pi r_h}} ~ \frac{a}{a_{rec}}~,}
\eqn\ImpactB{ \rho_0 \simeq \rho_c \sqrt{\frac{a_{min}}{3a}} = \rho_c \sqrt{\frac{a_{rec}}{a}} \left( \frac{\pi}{32}\frac{r_h}{\rho_c} \right)^{1/4}
\ll \rho_c~,}
and only the first solution obeys the condition \bbbR\  characterizing a configuration with minimal energy. Thus, for given values of $(a,\rho_1,\rho_2,r_h)$ satisfying \ThetaBrec, equation \ImpactA\  characterizes (through \aaaT) the profile of a connected $D4$-brane which corresponds to a global minimum of energy.  The profile with $\rho_0$ given by \ImpactB\  describes a local maximum of energy which separates the global minimum from the configuration of straight disconnected branes, which corresponds to a local minimum of the energy.  As $a$ decreases, the trajectories characterized  by \ImpactA\  and \ImpactB\  become closer and their energies change in opposite directions (see figure 7).  At $a=a_{rec}$ the minimum characterized by \ImpactA\  and the disconnected brane minimum have the same energy.  For $a_{min} < a < a_{rec}$, the connected minimum is a local minimum, while the disconnected configuration is the global minimum.   Finally, when $a$ decreases to $a_{min}$, the maximum and local minimum degenerate to the same profile, with $\rho_0 = \rho_c$, and disappear.

In the setting of section 4.2, one varies $r_h$ while keeping $a$ fixed.  To analyze this situation it is convenient to rewrite \ThetaBrec\  as
\eqn \Rrec{ r_h<(r_h)_{rec} \equiv \frac{\rho_1\rho_2a^2}{4(\rho_1+\rho_2)} \ll (r_h)_{max}~.}
We conclude that the reconnected brane (with $\rho_0$ given by \ImpactA) corresponds to the global minimum if $r_h<(r_h)_{rec}$, to a metastable state if $(r_h)_{rec}<r_h<(r_h)_{max}$ and it does not exist for $r_h>(r_h)_{max}$.

In the preceding discussion the endpoints of the $D4$-brane were assumed to have fixed locations, \EndPoint.  While it is clear that sliding along the $NS_{2}'$-brane increases the energy of the $D4$-brane, sliding in the ${\rm Re}(v)$-direction along the $NS$-brane seems energetically advantageous.  This is true, but it is easy to show that this effect leads to a subleading contribution to the energy.

We conclude this appendix by rewriting expressions \RhMax\ and \Rrec\ in terms of coordinates of the branes.  To leading order in $a$ and $y_i/y_{NS}$, we find
\eqn\Smmry{ \eqalign{ (r_h)_{max} & =  \left(\frac{16a^3}{27\pi}\right)^{1/2} \frac{(y_1-y_{NS})(y_2-y_1)}{y_2-y_{NS}} \simeq \frac{4 a^{3/2}}{3 \sqrt{3\pi}} \Delta y~, \cr
(r_h)_{rec} & =  \frac{a^2(y_1-y_{NS})(y_2-y_1)}{4(y_2-y_{NS})} \simeq \frac{a^2 \Delta y}{4}~.}}
%


%

\listrefs
\end